\documentclass[conference]{IEEEtran}
\IEEEoverridecommandlockouts
\usepackage{cite}
\usepackage{amsmath,amssymb,amsfonts}
\usepackage{algorithmic}
\usepackage{graphicx}
\usepackage{textcomp}
\usepackage{xcolor}
\def\BibTeX{{\rm B\kern-.05em{\sc i\kern-.025em b}\kern-.08em
    T\kern-.1667em\lower.7ex\hbox{E}\kern-.125emX}}
    
\usepackage{multirow}
\usepackage[utf8]{inputenc}

\begin{document}

%\title{Hash Code Indexing in Cross-Modal Retrieval\\
%\title{Efficient Indexing in Cross-Modal Hashing-Based Datasets \\
%\title{Index Learning in Cross-Modal Hashing-Based Datasets\\

\title{Effective and Efficient Indexing in Cross-Modal Hashing-Based Datasets\\
{\footnotesize \textsuperscript{}}
\thanks{}
}

\author{\IEEEauthorblockN{Sarawut Markchit}
\IEEEauthorblockA{\textit{Computer Science and Information Engineering} \\
\textit{National Chiayi University}\\
Chiayi, Taiwan, R.O.C. \\
s1040465@mail.ncyu.edu.tw}
\and
\IEEEauthorblockN{Chih-Yi Chiu}
\IEEEauthorblockA{\textit{Computer Science and Information Engineering} \\
\textit{National Chiayi University}\\
Chiayi, Taiwan, R.O.C. \\
cychiu@mail.ncyu.edu.tw}
}

\maketitle

\begin{abstract}

To overcome the barrier of storage and computation, the hashing technique has been widely used for nearest neighbor search in multimedia retrieval applications recently.
Particularly, cross-modal retrieval that searches across different modalities becomes an active but challenging problem.
Although dozens of cross-modal hashing algorithms are proposed to yield compact binary codes, the exhaustive search is impractical for the real-time purpose, and Hamming distance computation suffers inaccurate results.
In this paper, we propose a novel search method that utilizes a probability-based index scheme over binary hash codes in cross-modal retrieval.
The proposed hash code indexing scheme exploits a few binary bits of the hash code as the index code.
We construct an inverted index table based on index codes and train a neural network to improve the indexing accuracy and efficiency.
Experiments are performed on two benchmark datasets for retrieval across image and text modalities, where hash codes are generated by three cross-modal hashing methods.
Results show the proposed method effectively boost the performance on these hash methods.
\end{abstract}

\begin{IEEEkeywords}
cross-modal hashing, inverted indexing, nearest neighbor search
\end{IEEEkeywords}

\section{Introduction}

Nearest neighbor (NN) search plays a fundamental role in machine learning and information retrieval.
Cross-modal retrieval, an application based on nearest neighbor search, has grabbed much research attention recently.
It is natural that multimedia data have multiple modalities; these modalities may contribute correlated semantic information, such as video-tag pairs in YouTube and image-text pairs in Flickr.
Cross-modal retrieval can return relevant results of one modality for a given query of another modality.
For example, we can use text queries to retrieve images, and use image queries to retrieve texts.
This retrieval paradigm provides a useful interface for users to search data across different modalities.

With the rapid growth of multimedia data, it is impractical to apply exhaustive search that consumes a tremendous computation resource in a large-scale dataset.
To address this issue, existing cross-modal retrieval methods mainly leverage the hashing technique to generate compact data representations.
The goal of hashing is to embed the data points from the original space into a Hamming space as binary hash codes.
It generally exploits inter/intra class correlations or underlying data distribution/manifold to learn a set of hash functions, so that similar binary codes are generated for similar data points.
Hamming distance computation between binary codes enables a fast nearest neighbor search through hardware-supported bit operations with least memory consumption.
However, the use of hashing has to face the critical problem of quantization loss after binary embedding.
Even though the hashing algorithms have proposed various learning strategies to reduce the loss, there exists an inevitable large information gap between a real-valued vector and the corresponding binary code.
Searching nearest neighbors in the binary Hamming space is therefore less accurate than that in the real-valued Euclidean space.

In this paper, we propose to utilize a novel index scheme over binary hash codes for cross-modal retrieval.
The proposed index scheme exploits a few binary bits of the hash code as the index code.
An index structure is built by compiling reference data points with the same index codes into lists of an inverted table.
Given a query, we estimate the relevance of each index code that implicitly reflects the probability distribution of nearest neighbors (ground truth) for the query.
The estimation is realized by a prediction model that learns a nonlinear mapping between the query of one modality and the index space of another modality through deep learning.
Then we traverse the index table from the top rank index codes with the highest relevance scores to retrieve high quality candidates for further examination.
We evaluate the proposed index scheme adopted on three state-of-the-art cross-modal hashing algorithms in two widely-used benchmark datasets.
Experimental results show the proposed method can effectively improve the search performance, in terms of retrieval accuracy and computation time.
The proposed index scheme can be built upon any binary code datasets generated by hashing algorithms to derive the following benefits:

\begin{itemize}
    \item Based on the built index structure, the retrieval process can achieve sub-linear time complexity through inverted table lookup, compared with the exhaustive search that takes linear time complexity.
    \item Given a query, the learned prediction model is employed to estimate the relevance scores of the index codes for a precise ranking, rather than ranking by inaccurate Hamming distances.
\end{itemize}

The remainder of this paper is organized as follows.
In Section 2, we discuss the previous work about cross-modal retrieval.
Section 3 presents the proposed probability-based index scheme and search method.
Section 4 shows experimental results.
Conclusion remarks are given in Section 5.

\section{Related Work}

The hashing technique can be classified into three main categories: uni-modal hashing, multi-view hashing, and cross-modal hashing.
Uni-modal hashing derives binary hash codes from a single type of features.
The seminal work includes locality-sensitive hashing \cite{b1} and iterative quantization \cite{b2}.
%However, conventional uni-modal hashing methods cannot support multi-modal search as nearest neighbor cannot be directly computed across different modalities.
Multi-view hashing utilizes multiple types of features to learn better binary codes \cite{b3}\cite{b4}\cite{b5}.
%It requires all the modalities should be observed for all data points including query points and those in database.
%However, it is difficult to acquire all modalities of all data points.
Cross-modal hashing (CMH) aims to facilitate information retrieval across different modalities.
It usually embeds multiple heterogeneous data into a common latent space where the discriminability or similarity correlation is preserved.

Existing CMH algorithms can be further divided into unsupervised and supervised approaches.
Unsupervised cross-modal hashing algorithms basically employ the data distribution to learn hash functions without the label information.
For example, composite correlation quantization (CCQ)\cite{b6} uses correlation-maximal mappings to transform data from different modality types into an isomorphic latent space.
Unsupervised generative adversarial hashing\cite{b7} exploits generative adversarial networks to train a generative model and a discriminative model. A correlation graph is used to capture the underlying manifold structure across different modalities.
Fusion similarity hashing (FSH)\cite{b8} constructs an undirected asymmetric graph to model the fusion similarity among different modalities and embeds the fusion similarity across modalities into a common Hamming space.
On the other hand, supervised cross-modal hashing algorithms leverage the label information to assist the learning process.
For example, deep cross-modal hashing (DCMH)\cite{b9} learns hash functions for corresponding modalities through deep neural networks (DNN).
A cross-modal similarity matrix that is defined by class labels is employed to learn the hash functions, so the Hamming space can preserve the characteristics of the similarity matrix.
Semantics-preserving hashing (SePH)\cite{b10} transforms semantic affinities to a probability distribution and approximates it with hash codes by using kernel logistic regression.
Discrete latent semantic hashing (DLSH)\cite{b11} learns the latent semantic representations of different modalities and then projects them into the shared Hamming space.
Discrete latent factor model (DLFH)\cite{b12} utilizes the discrete latent factor to model the supervised information and adopts the maximum likelihood loss function without relaxation.
Deep discrete cross-modal hashing (DDCMH)\cite{b13} learns discrete nonlinear hash functions by preserving the intra-modality similarity at each hidden layer of the networks and the inter-modality similarity at the output layer of each individual network.
Semi-supervised cross-modal hashing by generative adversarial network (SCH-GAN)\cite{b14} employs the generative model to select margin examples of one modality from unlabeled data for a query of another modality, while the discriminative model tries to distinguish the generated examples and true positive examples with respect to the query.

\section{Hash Code Indexing}

\begin{figure*}[!htb]
  \centering
  \includegraphics[height=0.29\textheight,width=1.00\textwidth]{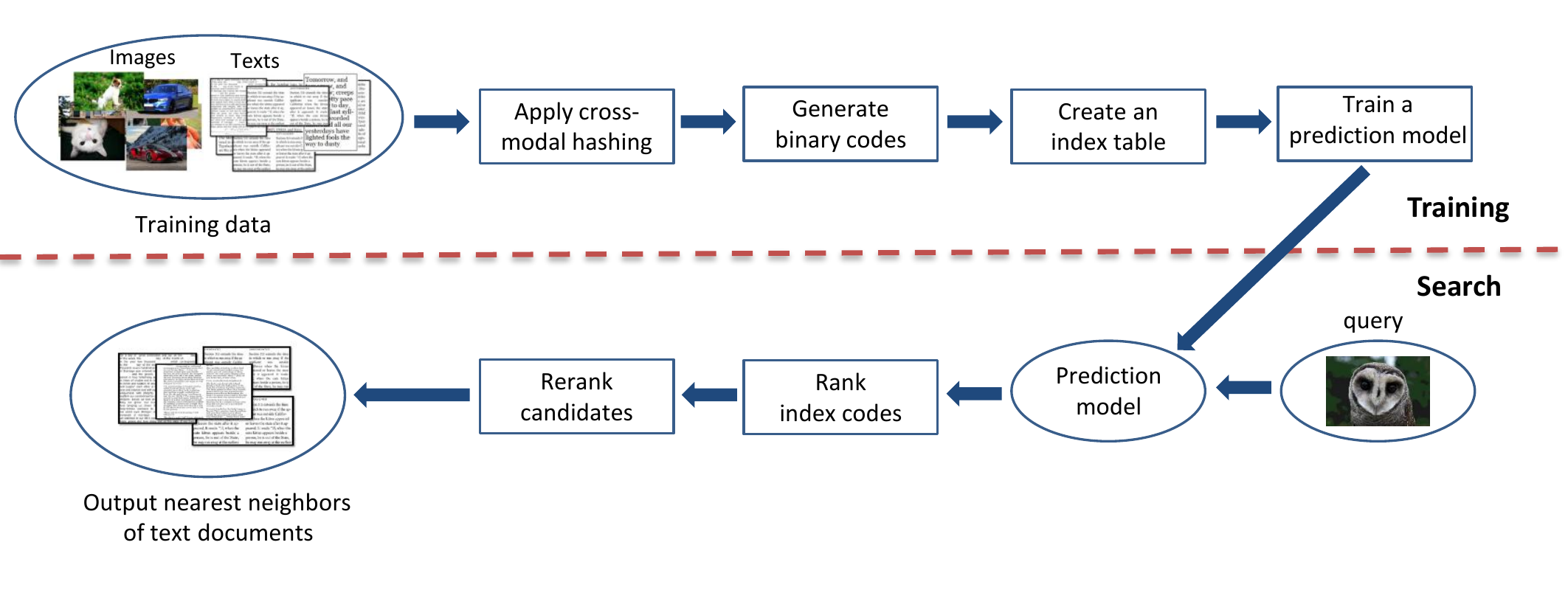} \caption{The proposed search framework of retrieving text documents for an image query.}
  \label{fig:framework}
\end{figure*}

Figure \ref{fig:framework} illustrates the proposed search framework with an example of retrieving text documents for an image query.
It consists of the training part and the search part.
In the training part, a CMH method is used to generate a reference dataset of binary codes of images and texts.
An inverted index table is created to organize the reference dataset. 
Then a prediction model is trained to estimate the relevance of each index code of the index table.
In the search part, the given image query is submitted to the prediction model to rank index codes based on their estimated relevance scores.
Candidates are retrieved from the top-rank index codes and reranked to output nearest neighbors of text documents in response to the query.
We elaborate the two parts in the following.

\subsection{Index Construction and Training}

Suppose that we have a reference dataset of \textit{N} binary codes of length \textit{c}, denoted as $B = \lbrace b_i \in \lbrace 0, 1 \rbrace ^c | i = 1, 2, ..., N \rbrace$.
The binary codes can be generated by any one of the CMH algorithms.
We select the first \textit{d} binary bits from $b_i$ as the index code $x_i \in \lbrace 0, 1 \rbrace ^d$.
%How to select the best \textit{d}-bit index code is related to the feature selection technique \cite{zhang2013binary}\cite{liu2017hash}, which is beyond the topic of this paper.
An index table with $2^d$ entries is constructed based on index codes, where each entry $E_\textbf{X}$ represents a particular index code \textbf{X} and attaches a set of associated reference data points:

  \begin{equation}
  E_\textbf{X} = \lbrace b_i | x_i = \textbf{X} \rbrace.
  \end{equation}

We train a prediction model that learns a nonlinear mapping between the query of one modality (e.g., texts) and the index space of another modality (e.g., images) through deep learning.
The model is used to estimate the relevance scores of index codes for a given query.
To compile the training dataset, we prepare a set of queries of one modality, denoted as $Q = \lbrace q_j^\theta | j = 1, 2, ..., J \rbrace$, where $q_j^\theta$ is the \textit{j}th query.
The relevant examples of another modality for $q_j^\theta$ are denoted as $\lbrace b_{jk}^\theta | k = 1, 2, ..., K \rbrace \in B$, where $b_{jk}^\theta$ is the \textit{k}th relevant example for $q_j^\theta$.
The definition of the relevant example is based on the class label information.
For example, the relevant examples of a text query are the images whose class is the same to the query.
The relevance score for each index code \textbf{X} is defined by the proportion of relevant examples to the entry size:
  \begin{equation}
  R_{j\textbf{X}}^\theta = \frac{\left | \lbrace b_{jk}^\theta | x_{jk}^\theta = \textbf{X} \rbrace \right |}{|E_\textbf{X}|},
  \end{equation}
where $| \cdot |$ denotes the set cardinality.
The training set is compiled as pairs of query features and relevance scores; the \textit{j}th query $q_j^\theta$ is associated with the set of $2^d$ relevance scores of index codes $\lbrace R_{j\textbf{X}}^\theta \rbrace$.

A fully-connected neural network is employed to learn the relation between the query and index codes based on the training set.
The input layer receives the feature representation of $q_j^\theta$, and the output layer predicts $2^d$ relevance scores of index codes $\lbrace P_{j\textbf{X}}^\theta \rbrace$.
Based on the cross-entropy loss between the predictions $\lbrace P_{j\textbf{X}}^\theta \rbrace$ and the target $\lbrace R_{j\textbf{X}}^\theta \rbrace$, we compute the error derivative with respect to the output of each neuron, which is backward propagated to each layer in order to update the weights of the neural network.

\subsection{NN Search}

Given a query \textit{q} for cross-modal retrieval, we utilize the trained network to predict the relevance scores of index codes $\lbrace P_\textbf{X} \rbrace$.
The index codes are ranked to select the top-\textit{R} index codes $\lbrace \textbf{X}_1, \textbf{X}_2, ..., \textbf{X}_R \rbrace$ with the highest relevance scores, and the reference data points associated with the top-ranking index codes are retrieved in a candidate set $C = \lbrace b_i | x_i \in \textbf{X}_r, r = 1, 2, ..., R \rbrace$.
We calculate the Hamming distance between the query and each of the candidates in \textit{C}, then sort the distances of the candidates in ascending order to return the desired number of NNs.

The time complexity for NN search mainly involves three parts, namely, the relevance score prediction, index code ranking, and candidate computation.
The time spent for relevance score prediction is related to the size of the neural network; it is regarded as a constant time.
Index code ranking requires to sort all index codes based on their relevance scores; it takes $2^d\cdot\log2^d = d\cdot2^d$ computation time.
Candidate computation is to compute the Hamming distances to the query for all candidates; it spends $s\cdot|C|$, where \textit{s} is a tiny constant time for computing the Hamming distance.
The candidate set \textit{C} is usually a fraction of the reference dataset \textit{B}, so we can reduce the computation time significantly compared with exhaustive search.
Interestingly, the quality of the candidate set is extremely good to further boost the search accuracy, as illustrated in the experimental section.

\section{Experiment}

To evaluate the proposed method, the experiment is conducted by using three state-of-the-art CMH algorithms on two widely-used benchmark datasets.
The benchmark datasets are MIRFlickr\cite{b15} and NUS-WIDE\cite{b16}, each of which consists of an image modality and a text modality.
Table \ref{tab:CMH_datasets} summarizes the properties of the two benchmark datasets, which are then used to produce the CMH datasets. 
The original MIRFlickr dataset has 25000 instances collected from the Flickr website. 
Each instance consists of an image, associated textual tag, and one or more of 24 predefined semantic labels. 
We removed textual tags that appear less than 20 times in the dataset, and then deleted instances that without any textual tag or semantic label. 
For each instance, its image view is characterized by a 150-D edge histogram, and its text view is represented as a 500-D feature vector derived from PCA on its binary tagging vector with respect to the textual tags. 
We took 5\% of MIRFlickr data to form the query set and the rest as the reference set. 
10000 instances were sampled from the reference set for training. 
The ground-truth neighbors were defined as those image-text pairs which share at least one common label.

For the original NUS-WIDE dataset, it has 260648 instances, each of which consists of an image and one or more of 81 predefined semantic labels. 
We selected 195834 image-text pairs that belong to the 21 most frequent concepts. 
The text for each point is represented as a 1000-dimensional bag-of-word vector. 
The hand-crafted feature for each image is a 500-dimensional bag-of-visual word (BOVW) vector. 
We used 2000 data points as the query set and the remaining points as the reference set. 20000 data points were sampled from the reference set for training. 
The ground truth neighbors were defined as those image-text pairs which share at least one common label, as the same to MIRFlickr.

The CMH algorithms, including SePH\cite{b10}, DCMH\cite{b9} and CCQ\cite{b6}, are employed to generate binary code datasets for MIRFlickr and NUS-WIDE.
The program was implemented in Python and run on a PC with Intel i7 CPU@3.6 GHz and 32GB RAM.

\begin{table}[!htb]
\centering
  \caption{CMH datasets}
  \label{tab:CMH_datasets}
  \begin{tabular}{|l|c|c|}
    \hline
    Dataset & MIRFlickr & NUS-WIDE\\
   \hline
	Reference set & 15902  & 193834 \\
	Training set & 10000  & 20000 \\
	Query set    & 836 & 2000 \\
	Number of Labels & 24 & 21 \\
	\hline
  \end{tabular}
\end{table}

\subsection{Implementation and Comparison}

For each CMH algorithm, three kinds of index schemes are implemented for comparison:

\begin{itemize}
    \item \textbf{Exhaustive.} It applies the exhaustive search that calculates Hamming distances between the query and all reference data without adopting any index structure.
    \item \textbf{Naïve-index (\textit{d} bits).} 
    It takes the first \textit{d} bits of the hash code as the index code for each reference data point. 
    The given query compares the index code to find candidates and then rerank the candidates according to their Hamming distances. Here $d = 14$.
    \item \textbf{DNN-index (\textit{d} bits).} It is the proposed method. In addition to the naïve index structure, we learn a 3-layer neural network to rank index codes. 
    The network is configured as I1-H2-H3-O4, where I1 is the input layer, H2 and H3 are hidden layers with the same number of units as I1, and O4 is the output layer. 
    ReLU and softmax are used as the activation functions for the hidden layers and output layer, respectively. Here $d \in \{8, 10, 12, 14\}$.
\end{itemize}

Mean average precision (MAP) is used to evaluate the retrieval accuracy for a set of queries $\textit{Q}$:
\begin{equation}
	\mathrm{MAP@R} = \frac{1}{|Q|} \sum_{i=1}^{|Q|} \frac{1}{R} 
    \sum_{j=1}^{R} pr(j) \cdot rel(j),
\end{equation}
where $R$ is the number of retrieved documents, $pr(j)$ denotes the precision of the top $j$ retrieved documents, and $rel(j) = 1$ if the $j$th retrieved document is relevant to the query, otherwise $rel(j) = 0$.
The relevant documents are defined as those image-text pairs which share at least one common label.
MAP is computed as the mean of all the queries’ average precision.

\begin{figure*}[!htb]
\centering
  \begin{tabular}{@{}cccc@{}}
    \includegraphics[height=.15\textheight,width=.25\textwidth]{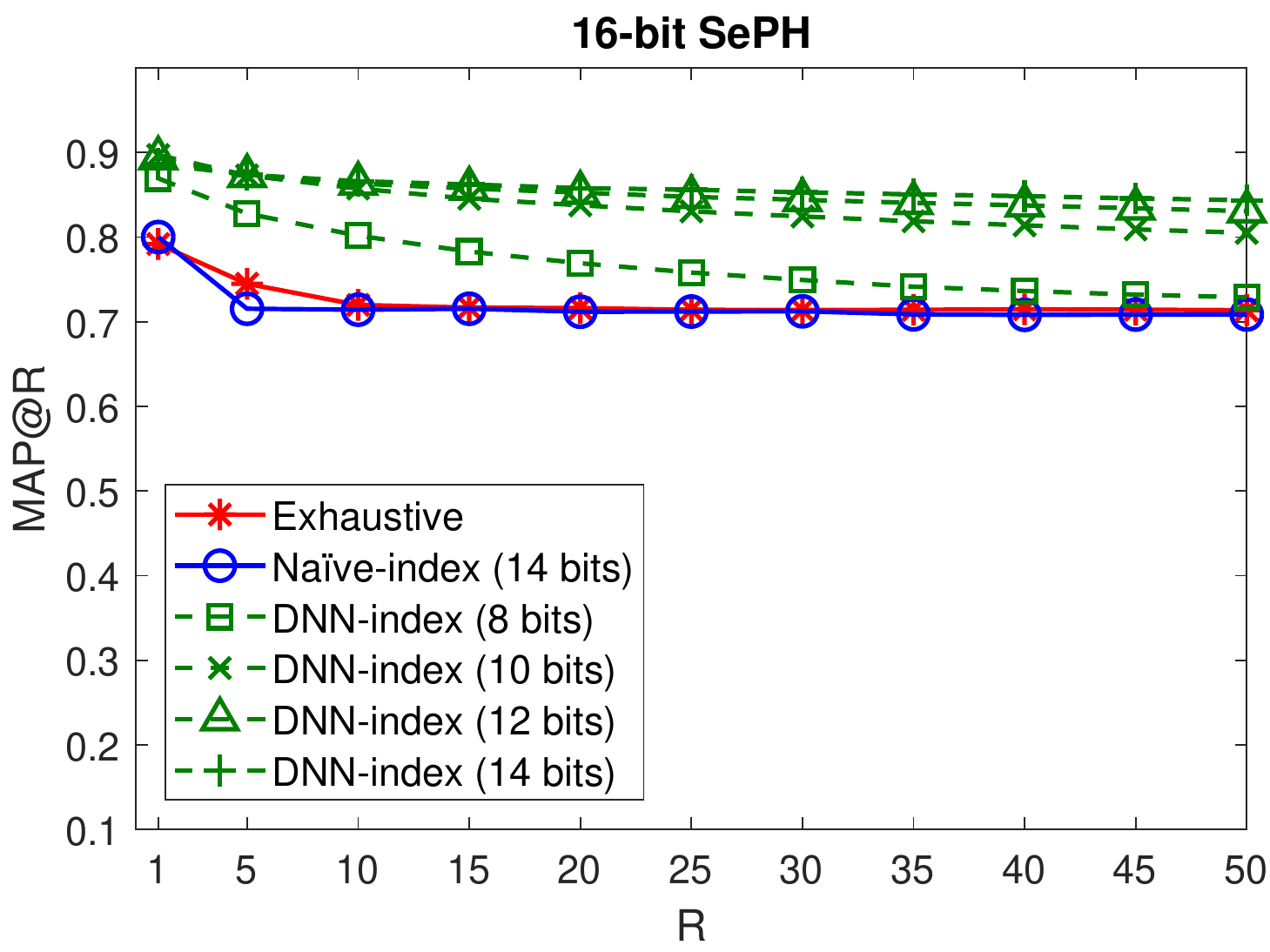} 
    \includegraphics[height=.15\textheight,width=.25\textwidth]{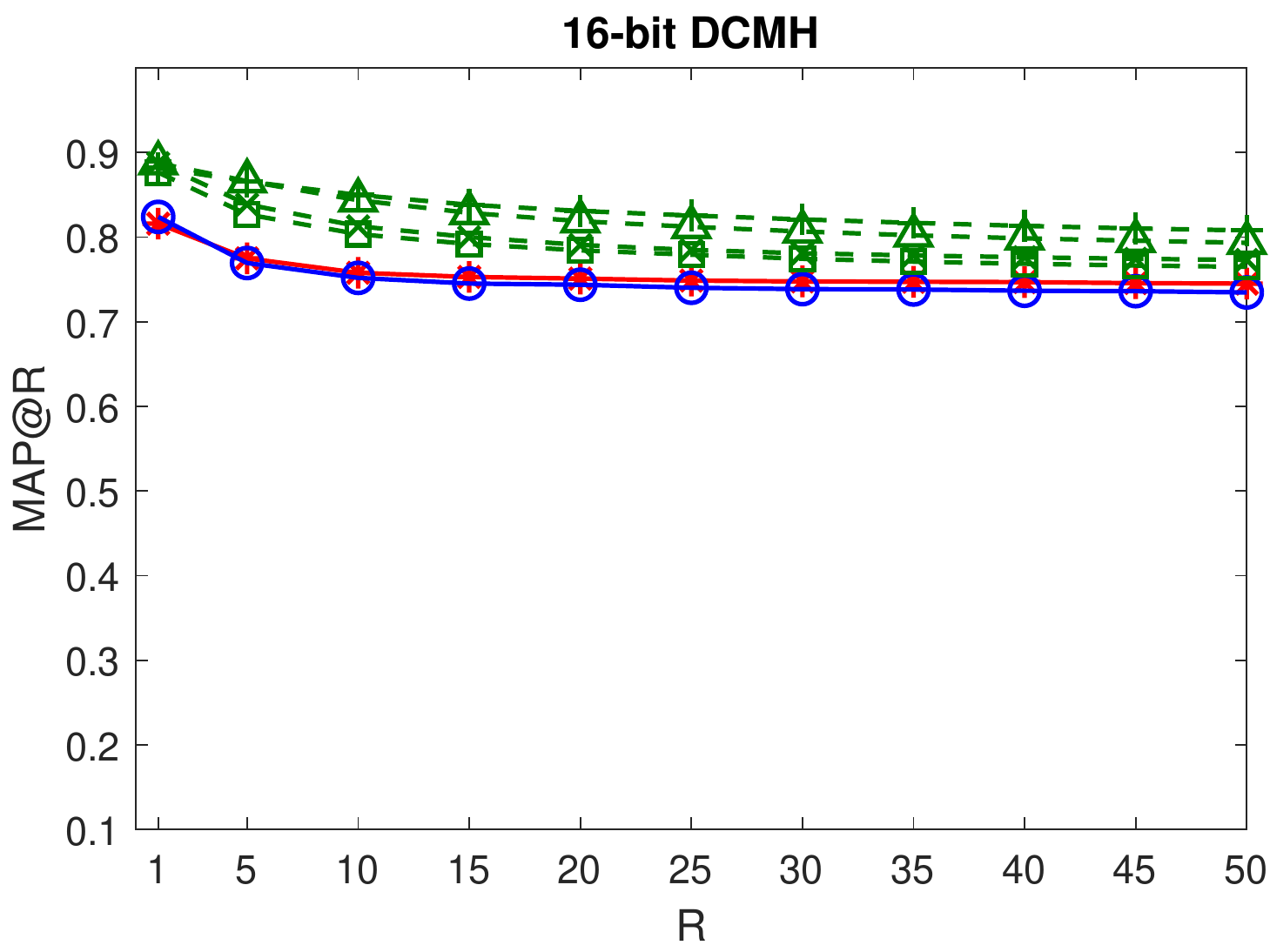} 
    \includegraphics[height=.15\textheight,width=.25\textwidth]{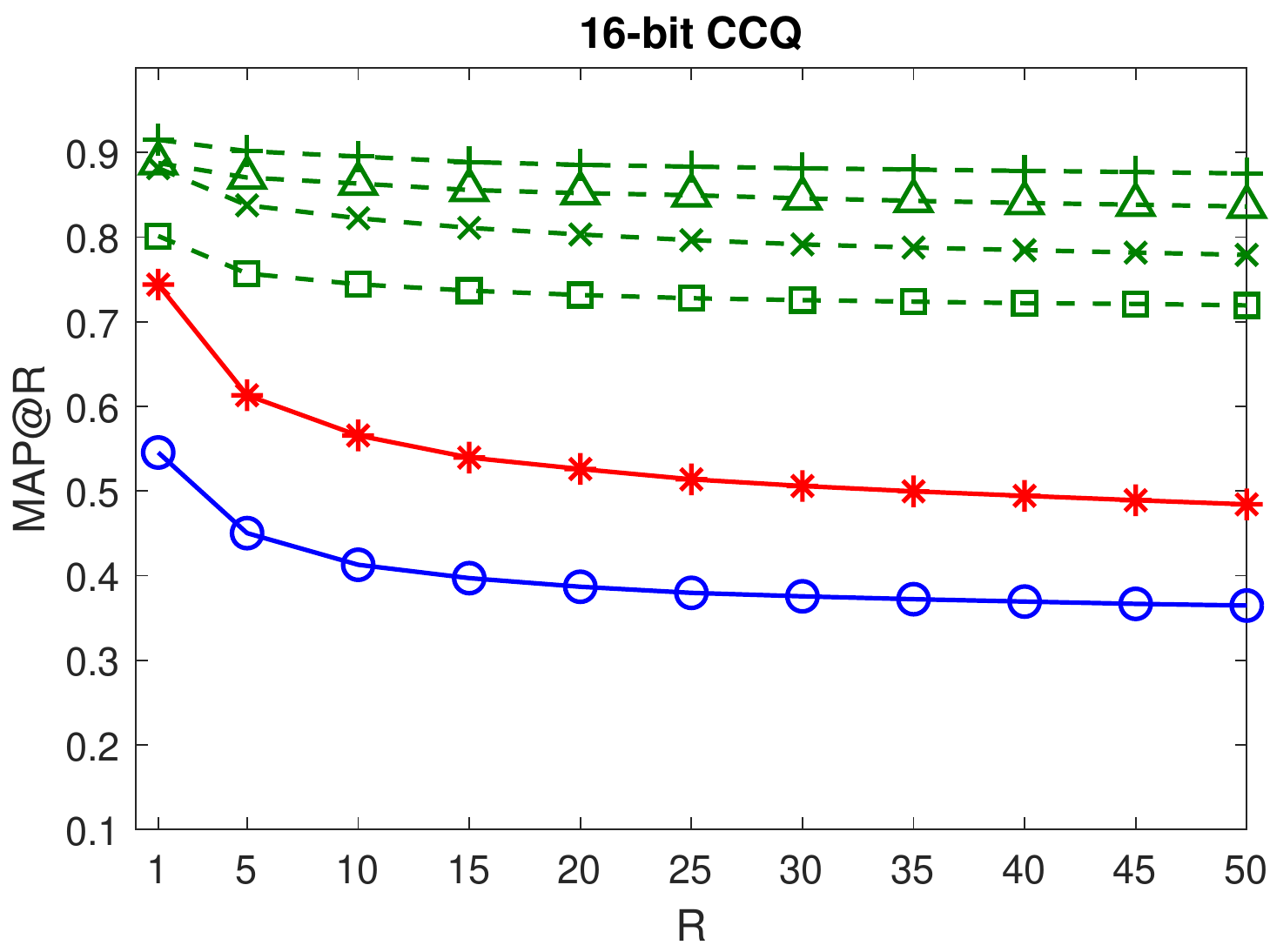}
    \\
    \includegraphics[height=.15\textheight,width=.25\textwidth]{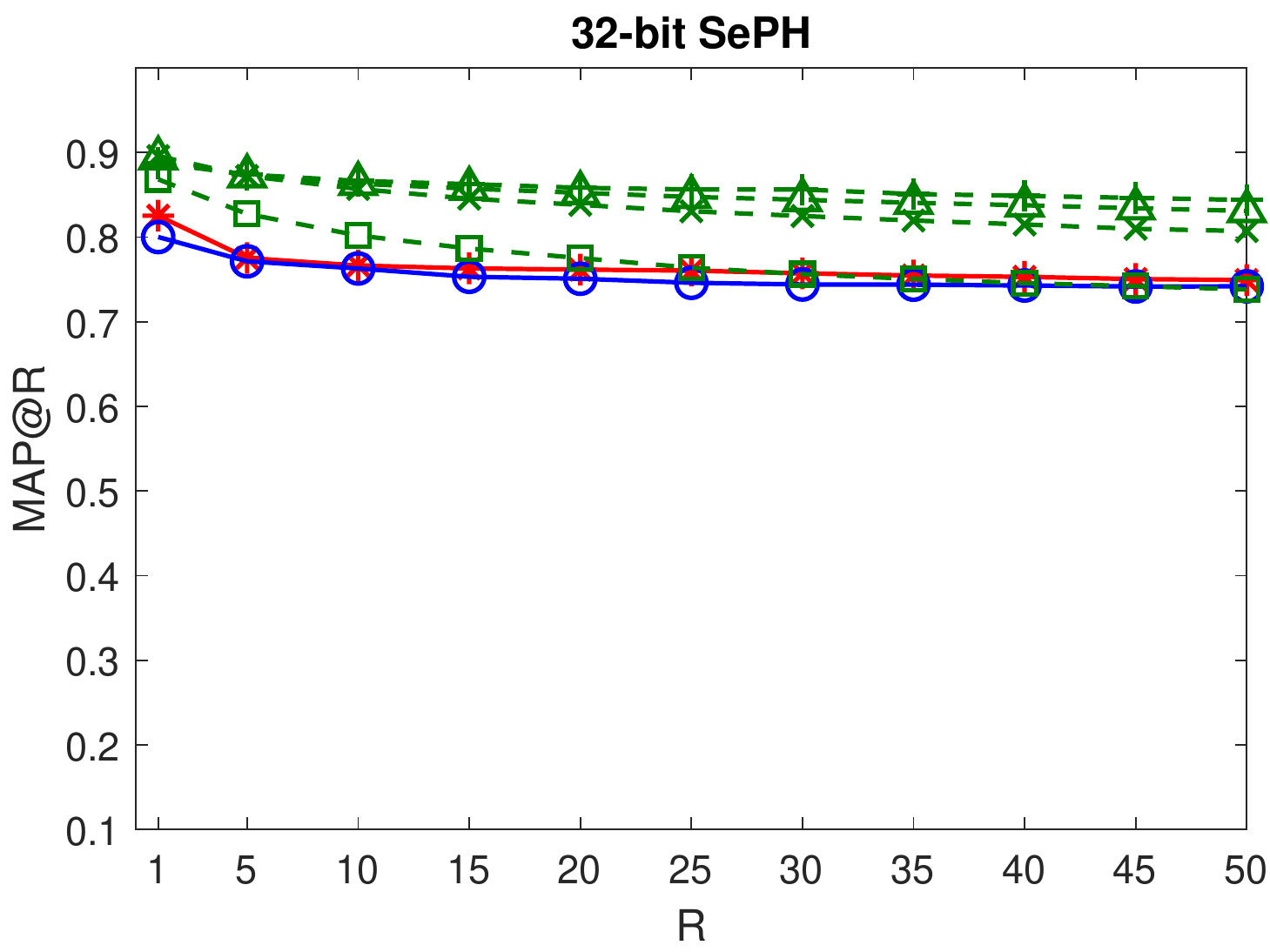}
    \includegraphics[height=.15\textheight,width=.25\textwidth]{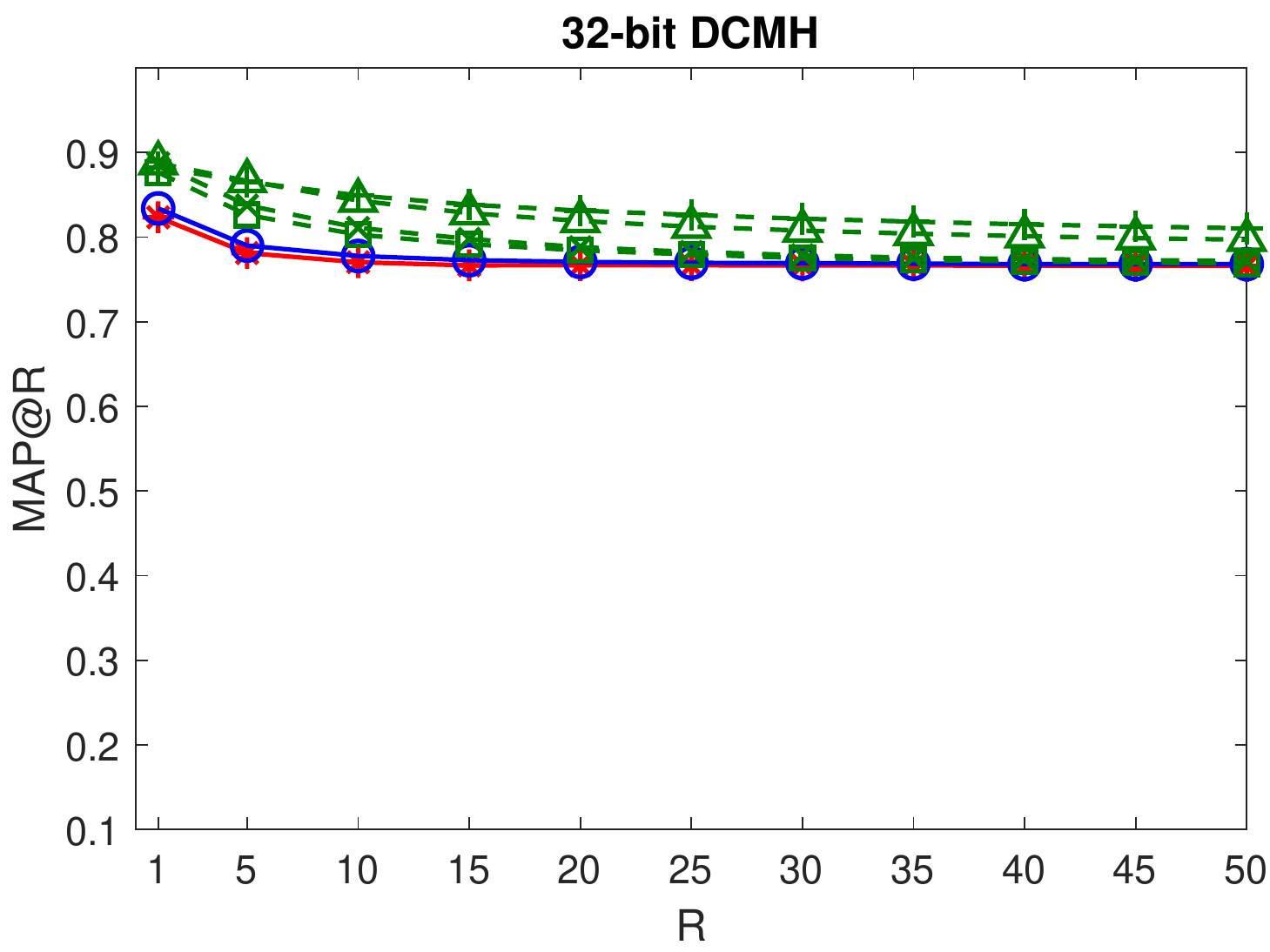} 
    \includegraphics[height=.15\textheight,width=.25\textwidth]{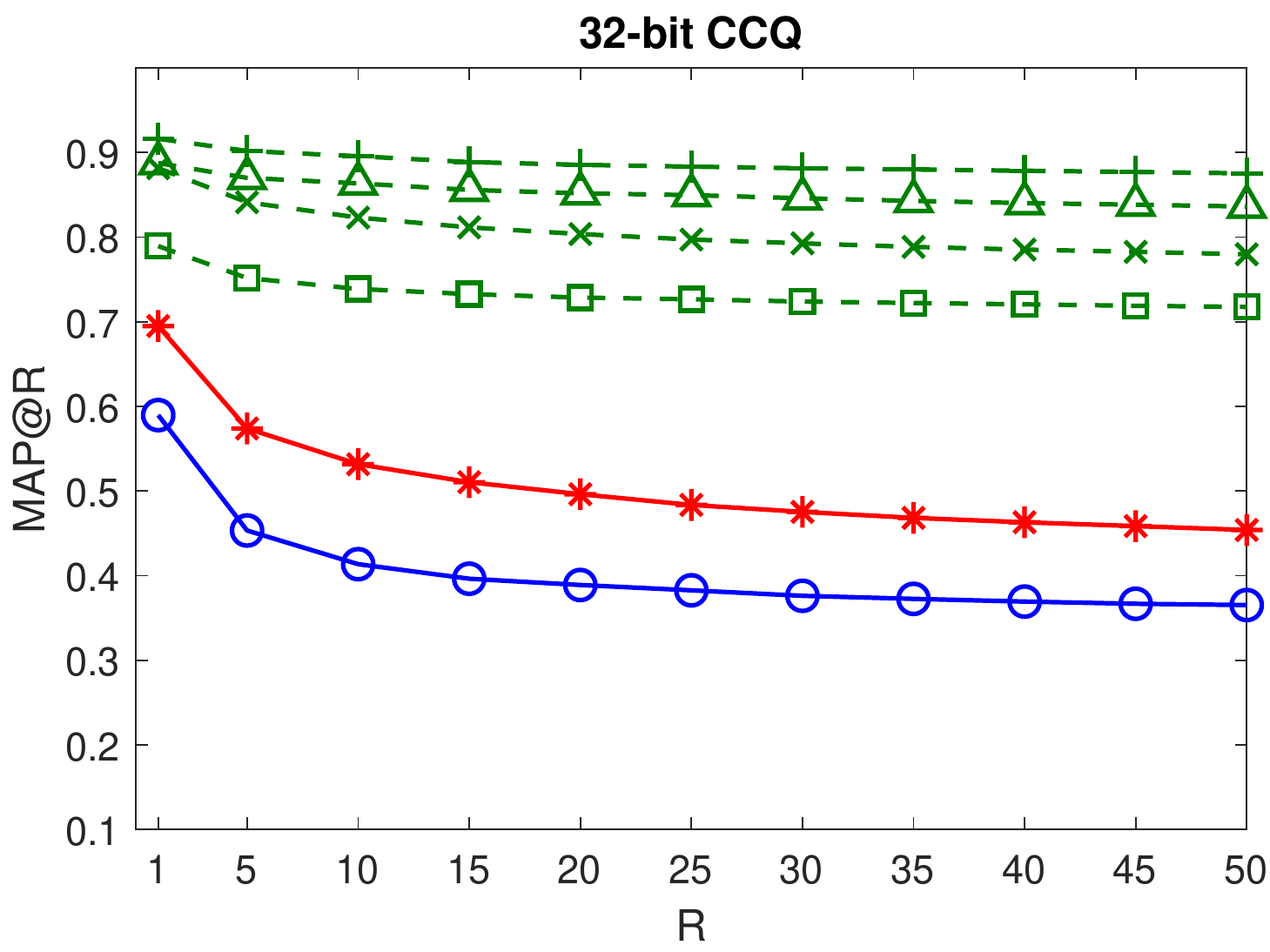} 
  \end{tabular}  
  \caption{MAP@R in the MIRFlickr dataset for text query vs. image dataset.}
  \label{fig:16_ti}
\end{figure*}

\begin{figure*}[!htb]
\centering
  \begin{tabular}{@{}cccc@{}}
    \includegraphics[height=.15\textheight,width=.25\textwidth]{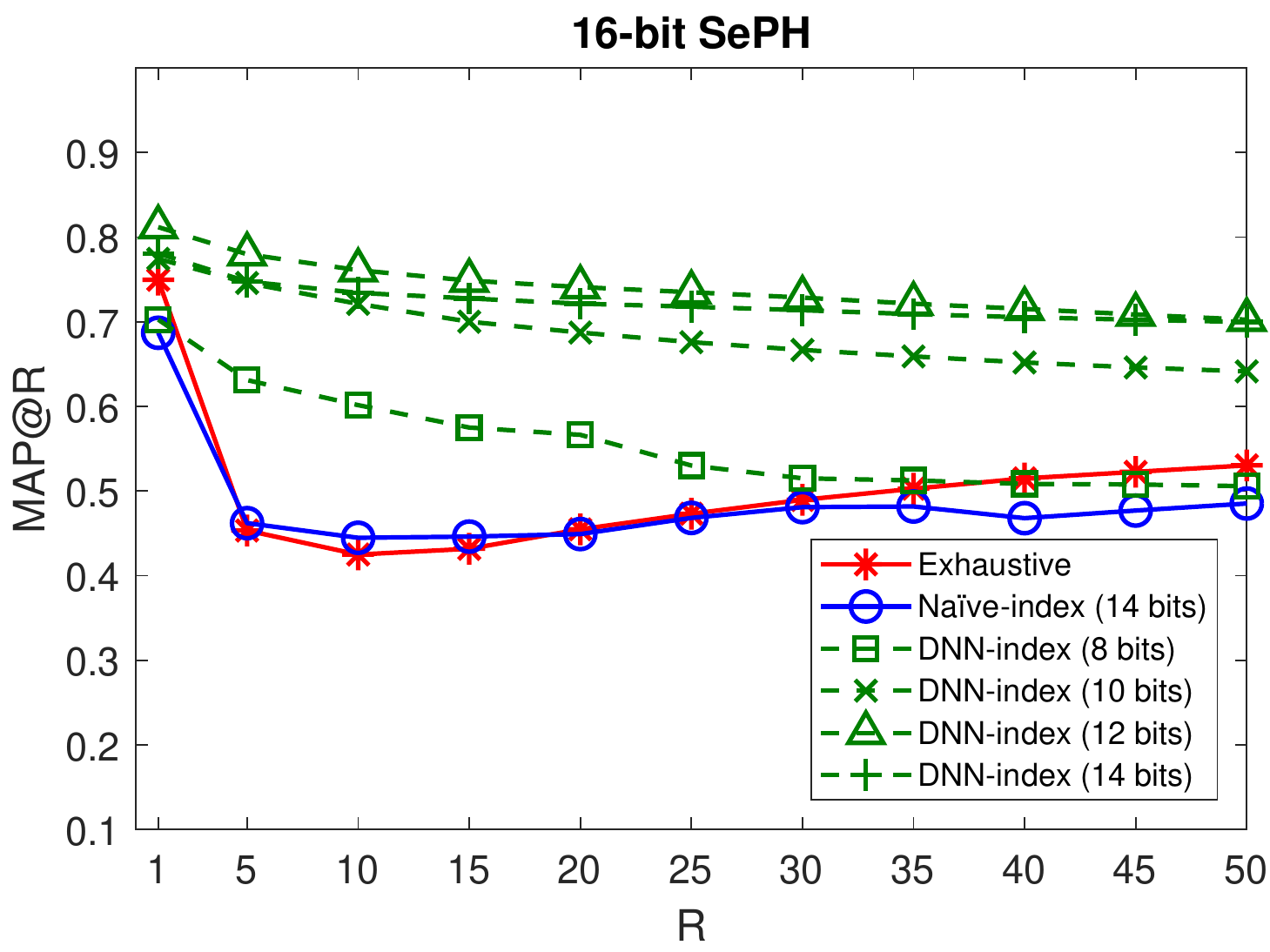} 
    \includegraphics[height=.15\textheight,width=.25\textwidth]{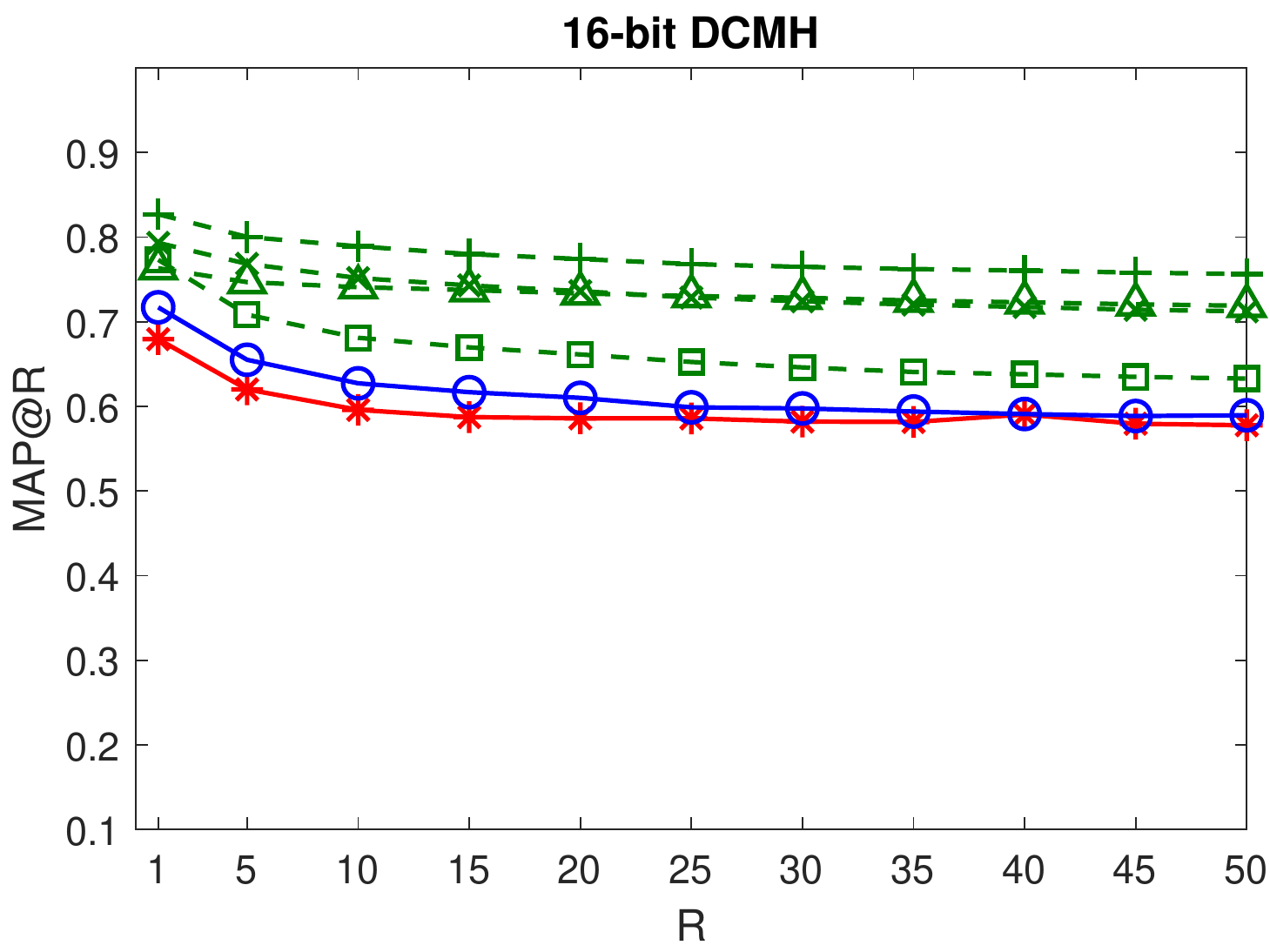} 
    \includegraphics[height=.15\textheight,width=.25\textwidth]{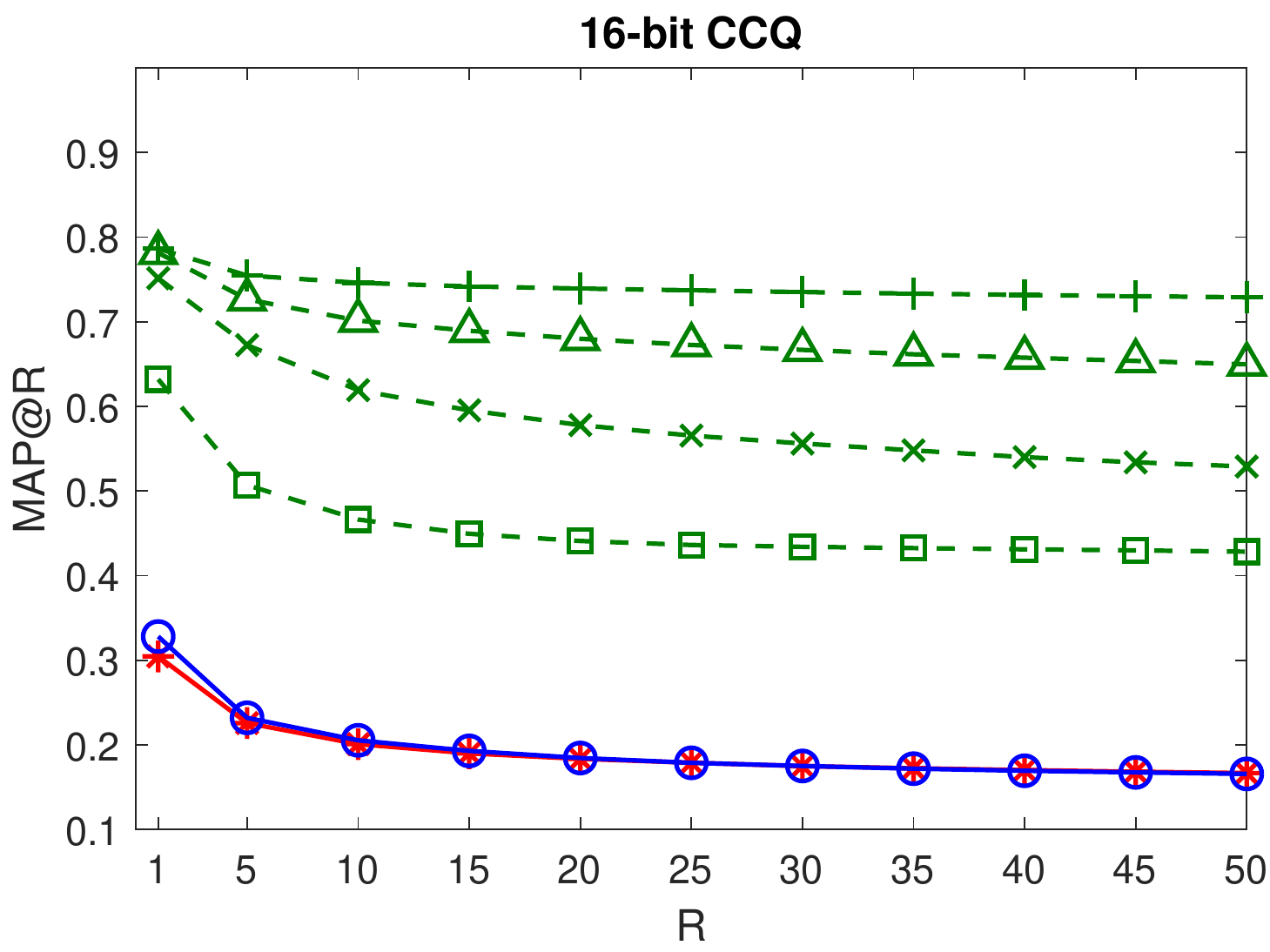}
    \\
    \includegraphics[height=.15\textheight,width=.25\textwidth]{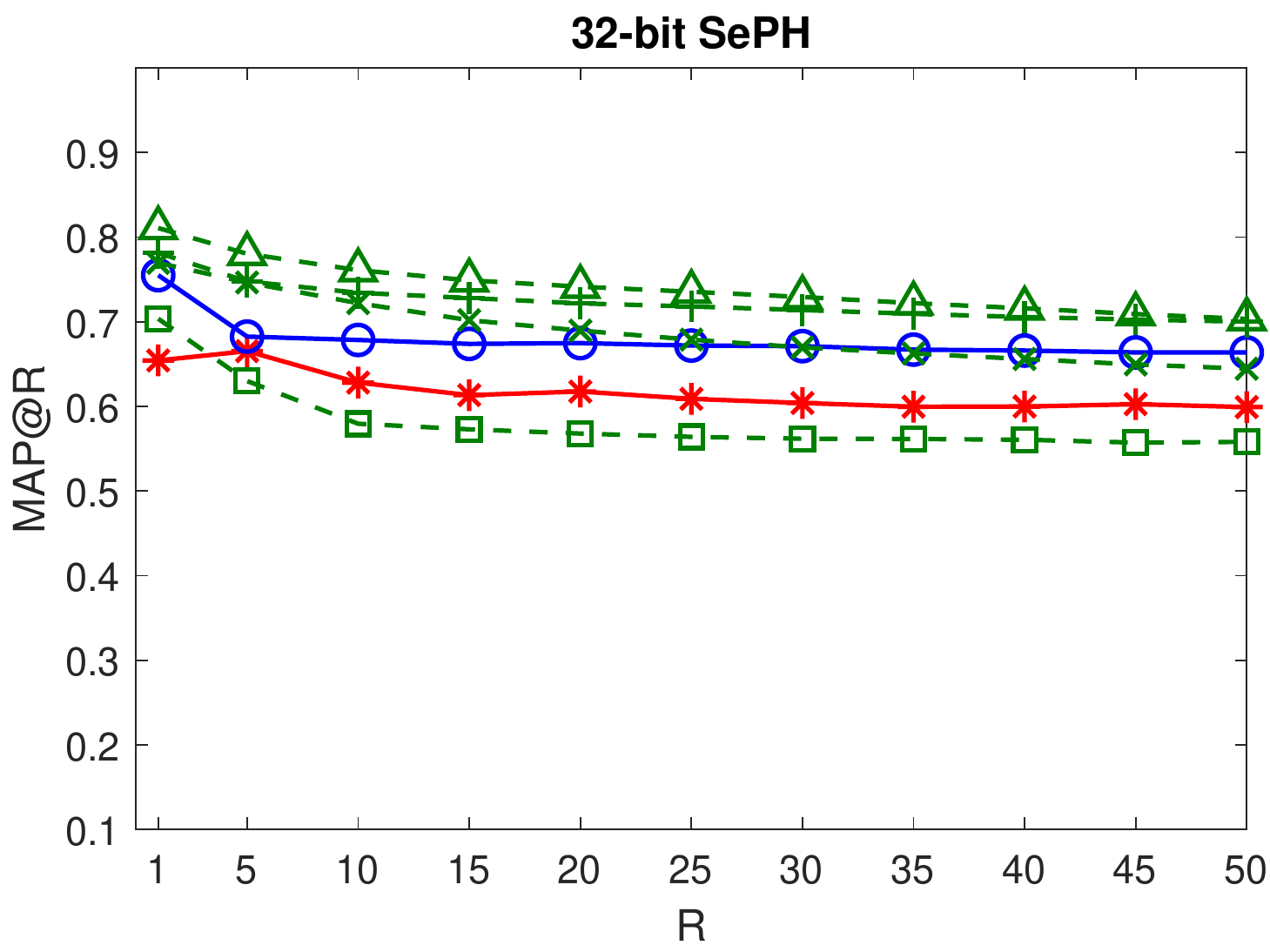} 
    \includegraphics[height=.15\textheight,width=.25\textwidth]{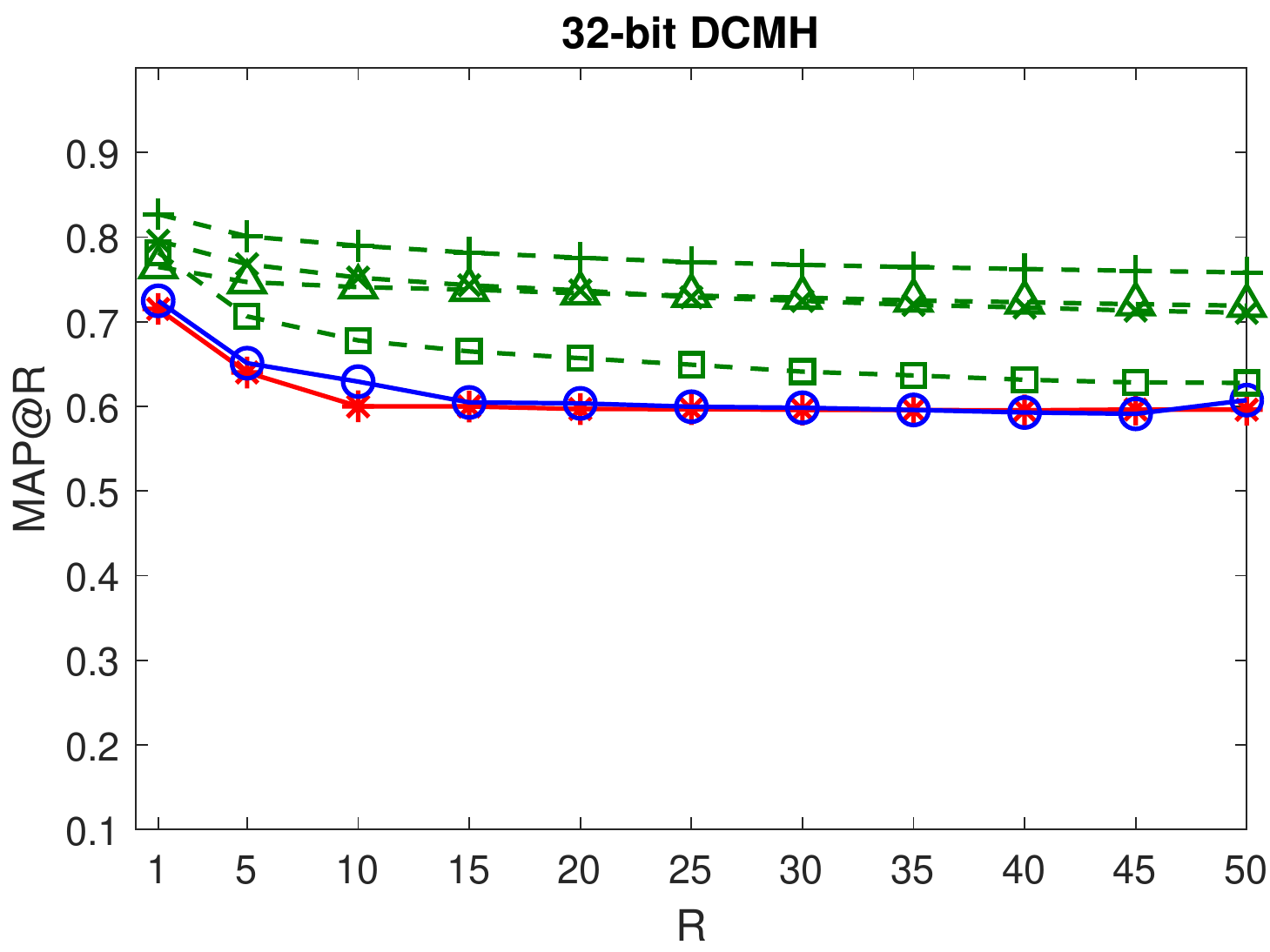} 
    \includegraphics[height=.15\textheight,width=.25\textwidth]{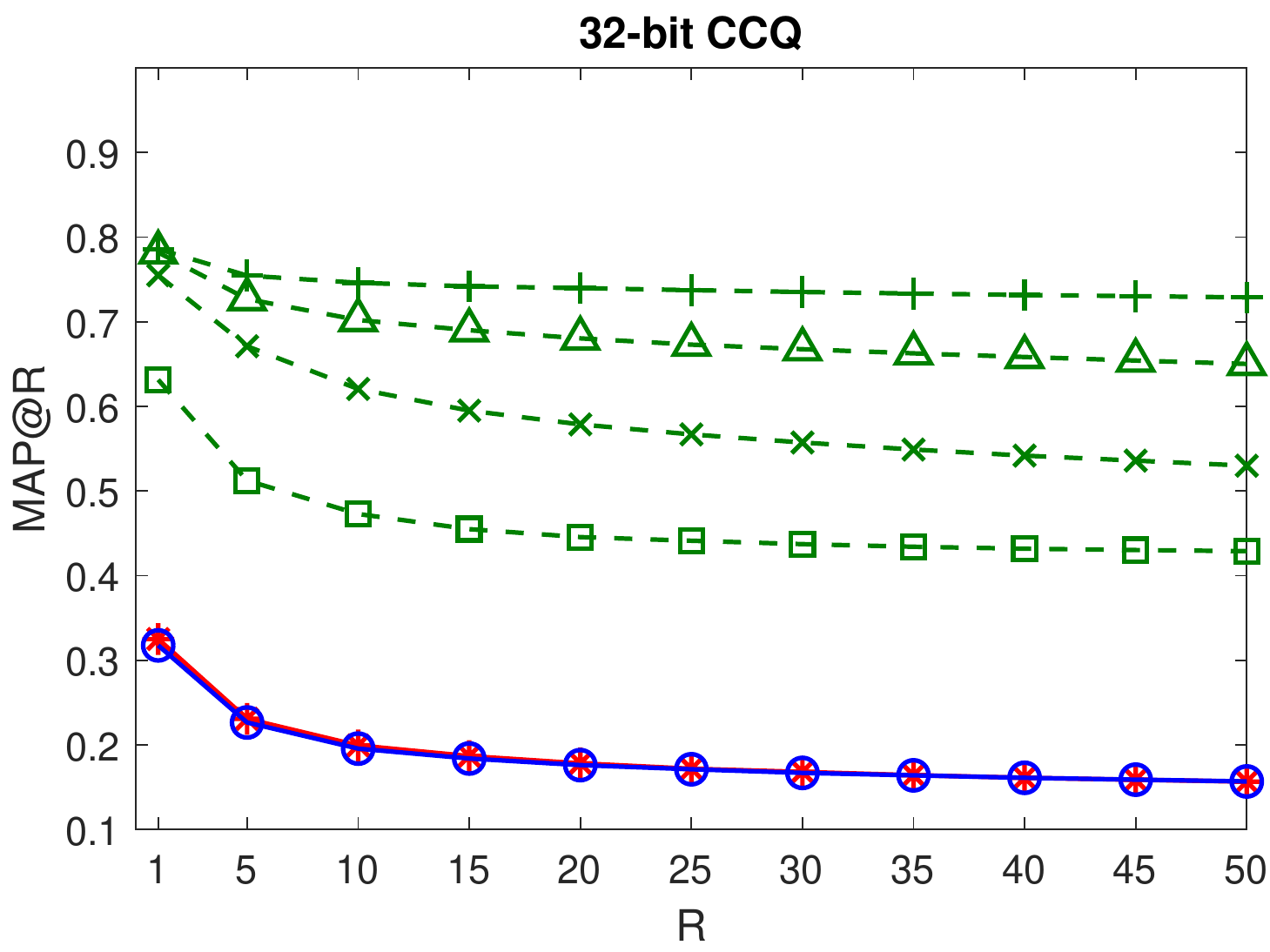} 
  \end{tabular}  
  \caption{MAP@R in the NUS-WIDE dataset for text query vs. image dataset.}
  \label{fig:32_ti}
\end{figure*}

\begin{figure*}[!htb]
\centering
  \begin{tabular}{@{}cccc@{}}
    \includegraphics[height=.15\textheight,width=.25\textwidth]{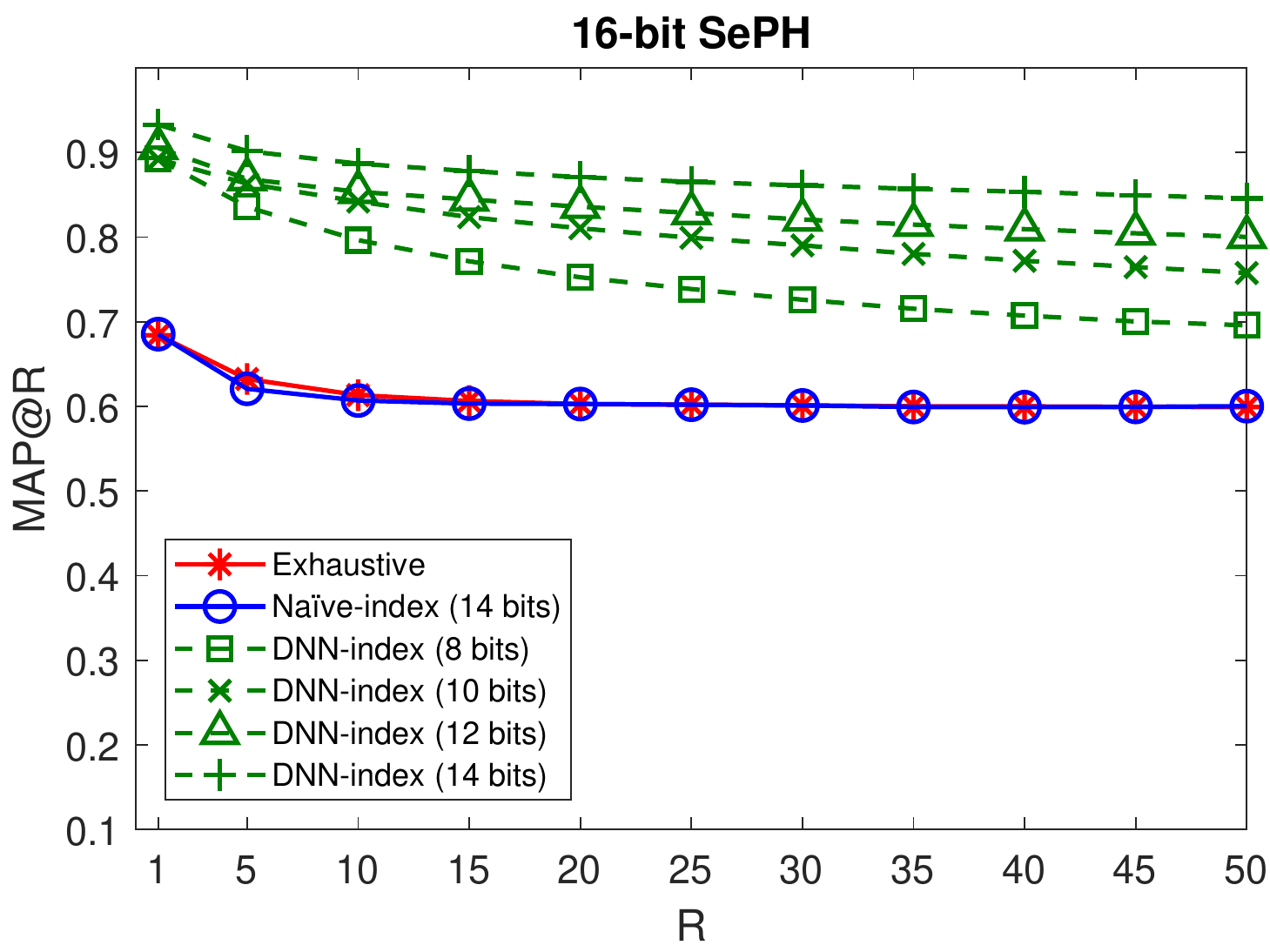} 
    \includegraphics[height=.15\textheight,width=.25\textwidth]{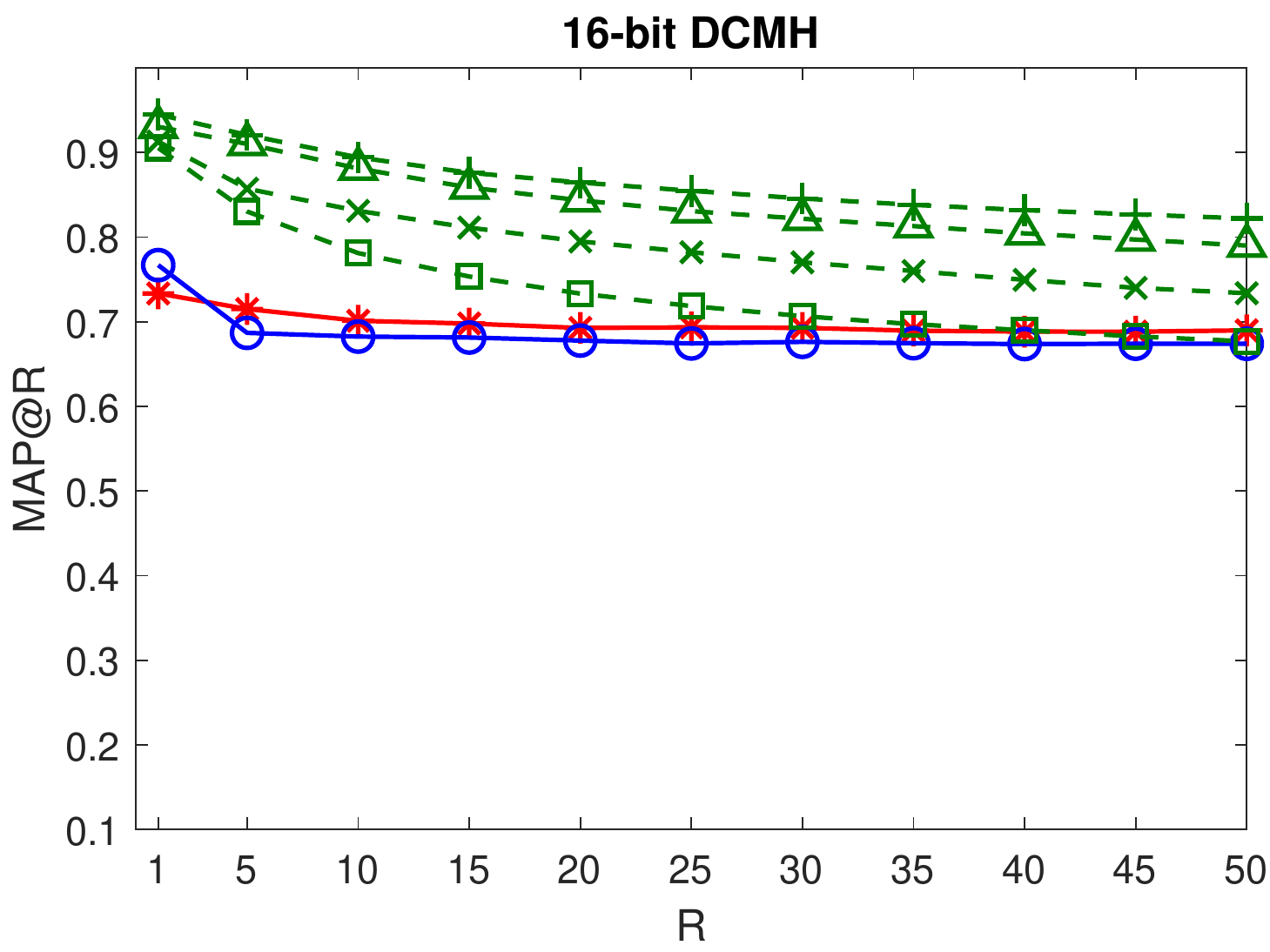} 
    \includegraphics[height=.15\textheight,width=.25\textwidth]{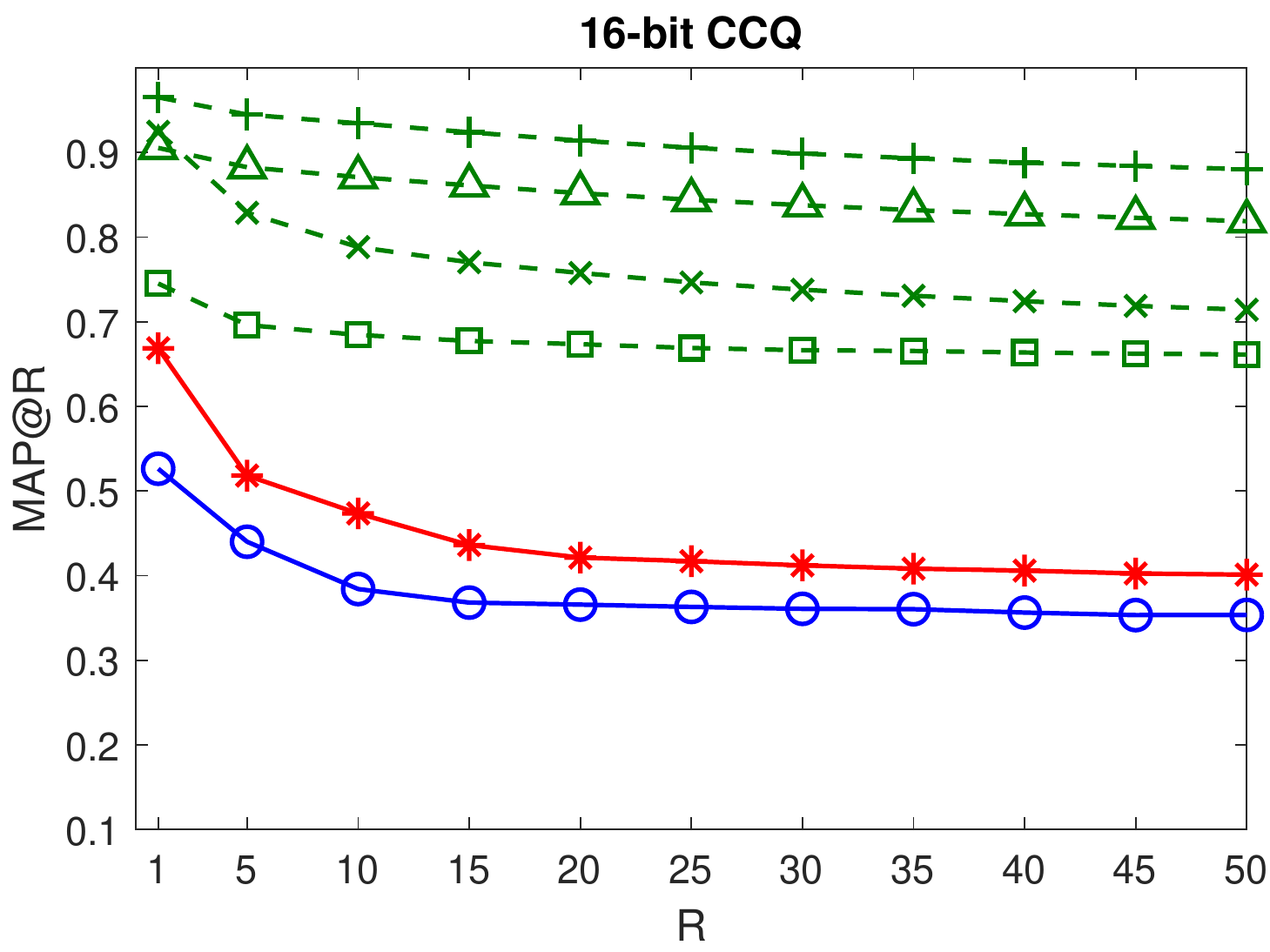}
    \\
    \includegraphics[height=.15\textheight,width=.25\textwidth]{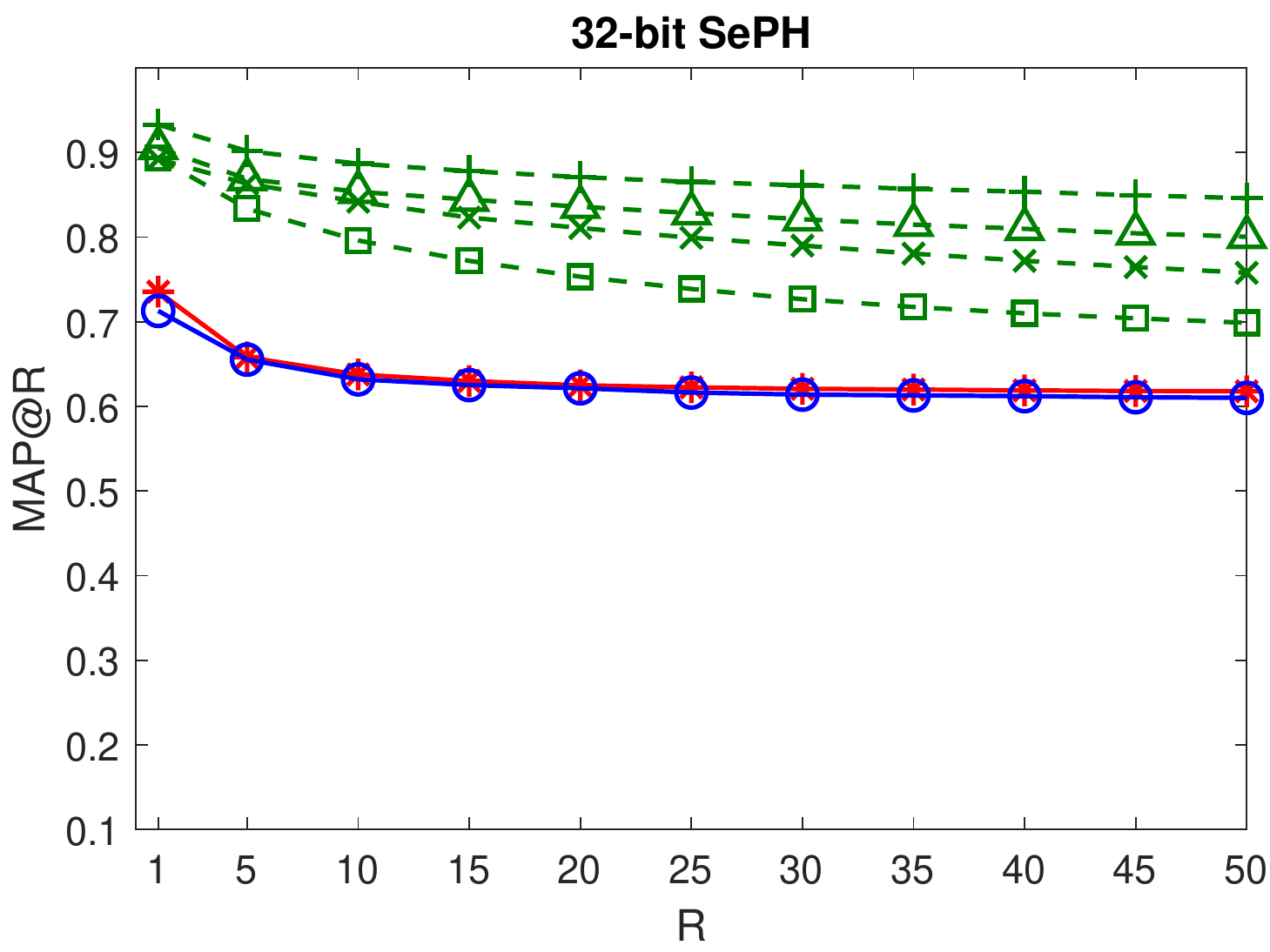}
    \includegraphics[height=.15\textheight,width=.25\textwidth]{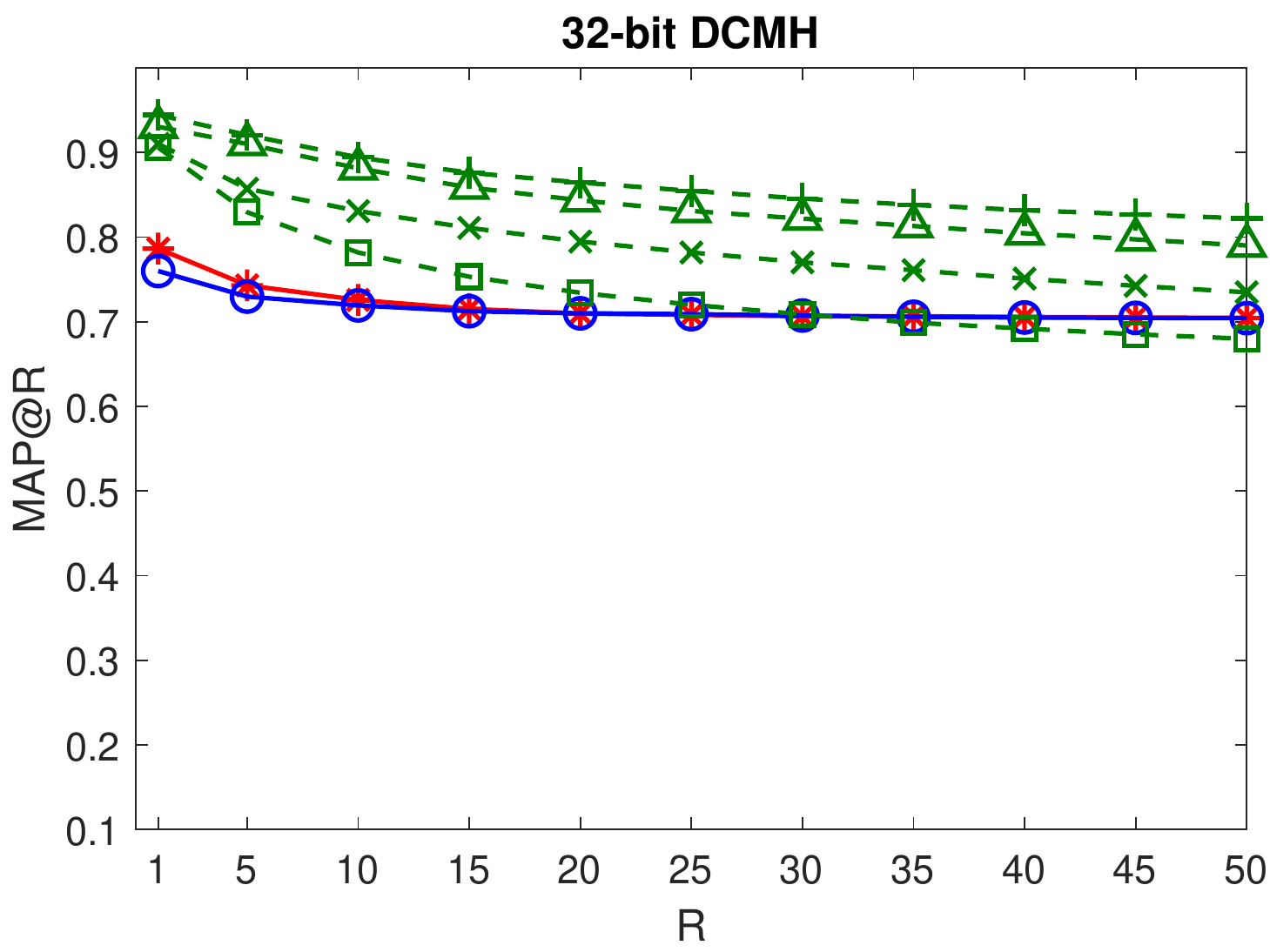} 
    \includegraphics[height=.15\textheight,width=.25\textwidth]{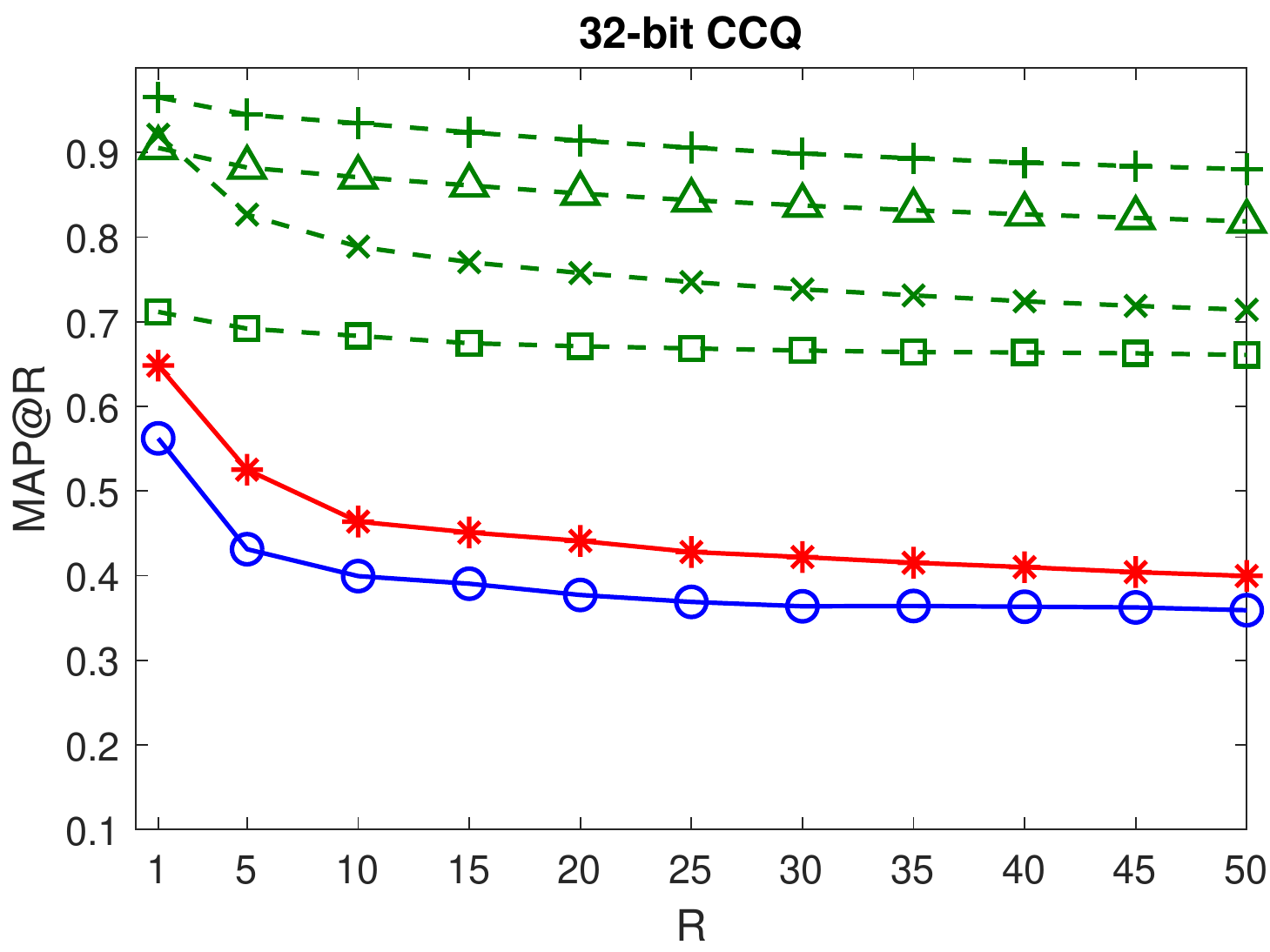} 
  \end{tabular}  
  \caption{MAP@R in the MIRFlickr dataset for image query vs. text dataset.}
  \label{fig:16_it}
\end{figure*}

\begin{figure*}[!htb]
\centering
  \begin{tabular}{@{}cccc@{}}
    \includegraphics[height=.15\textheight,width=.25\textwidth]{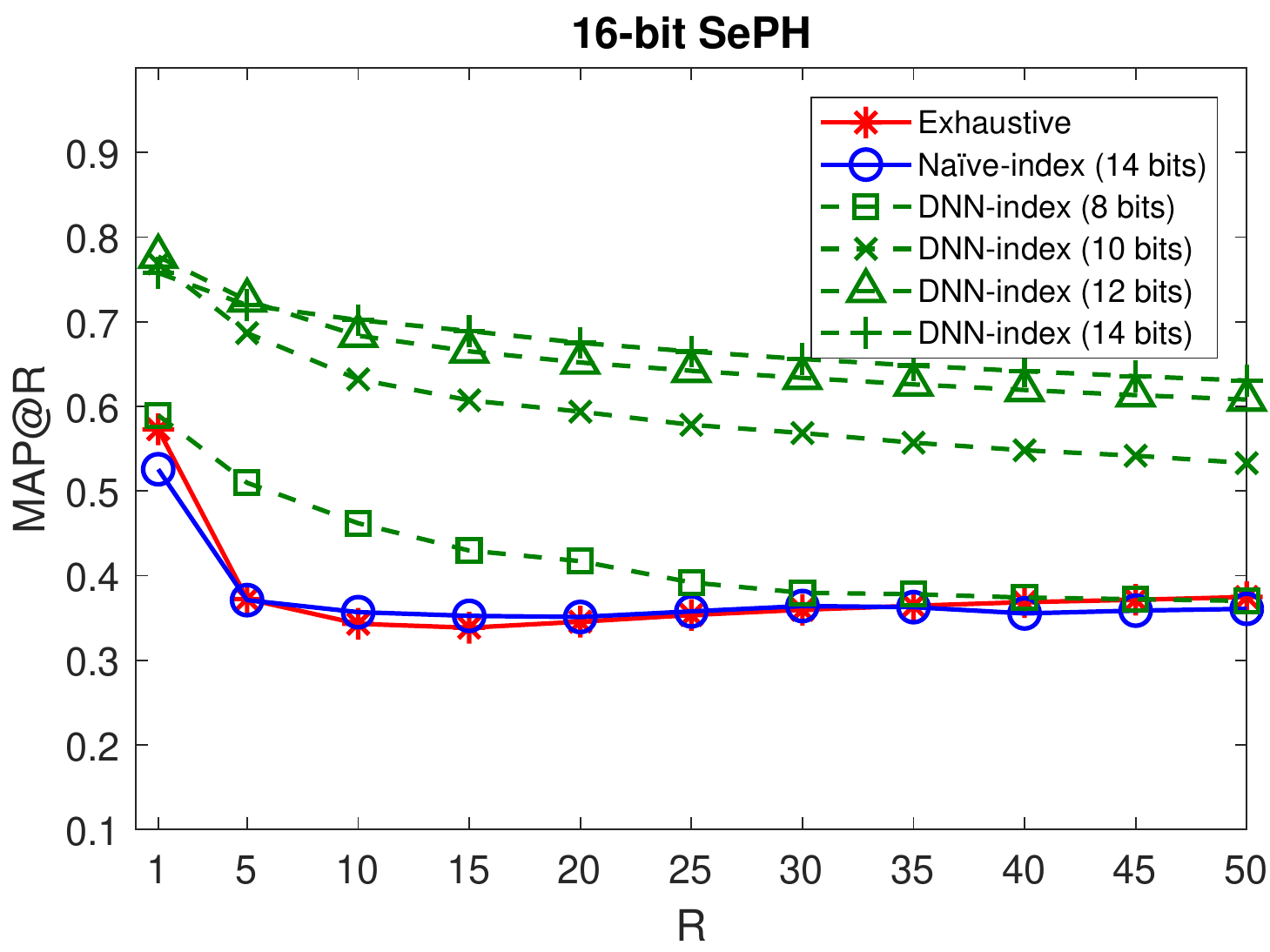} 
    \includegraphics[height=.15\textheight,width=.25\textwidth]{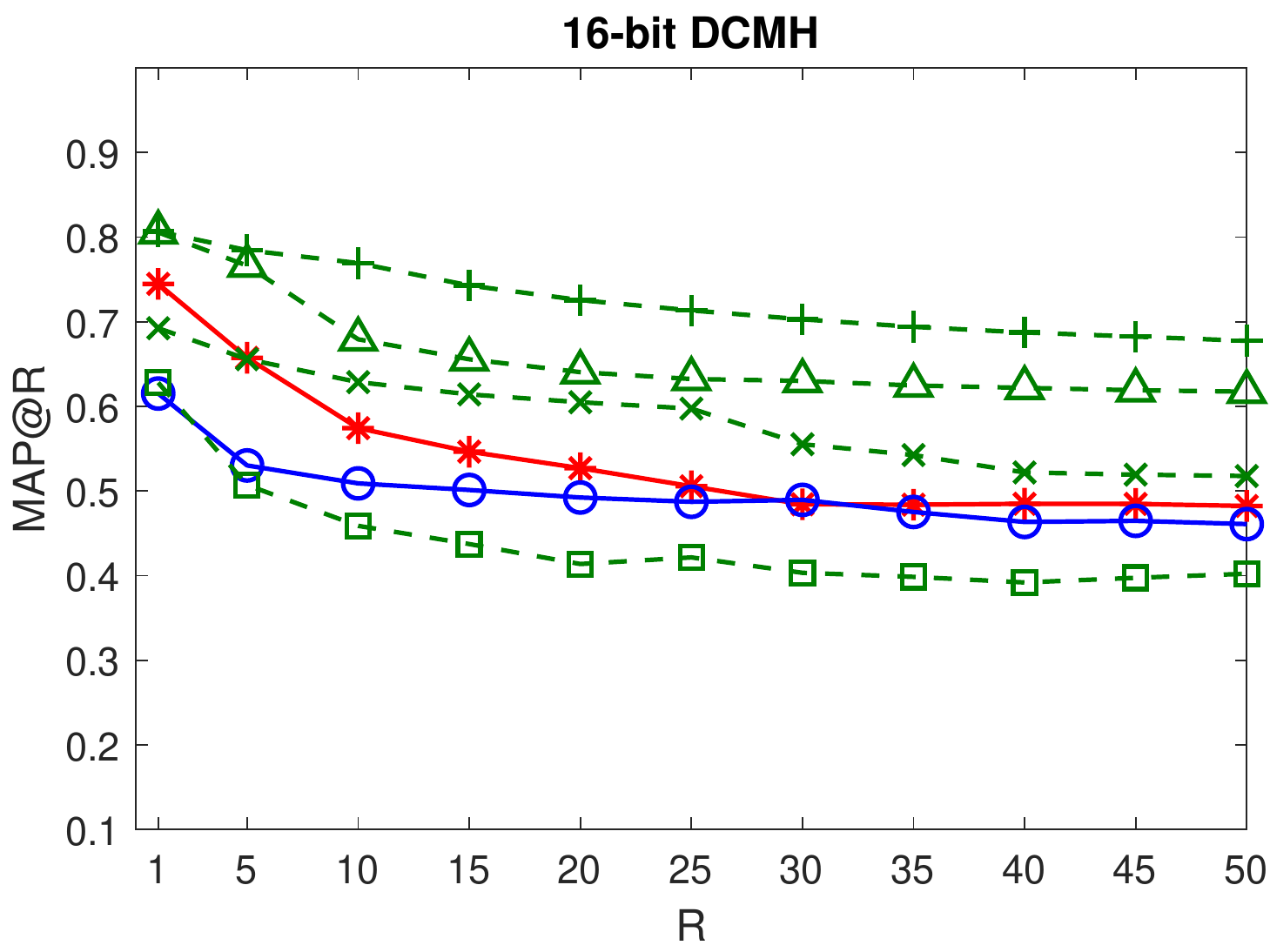} 
    \includegraphics[height=.15\textheight,width=.25\textwidth]{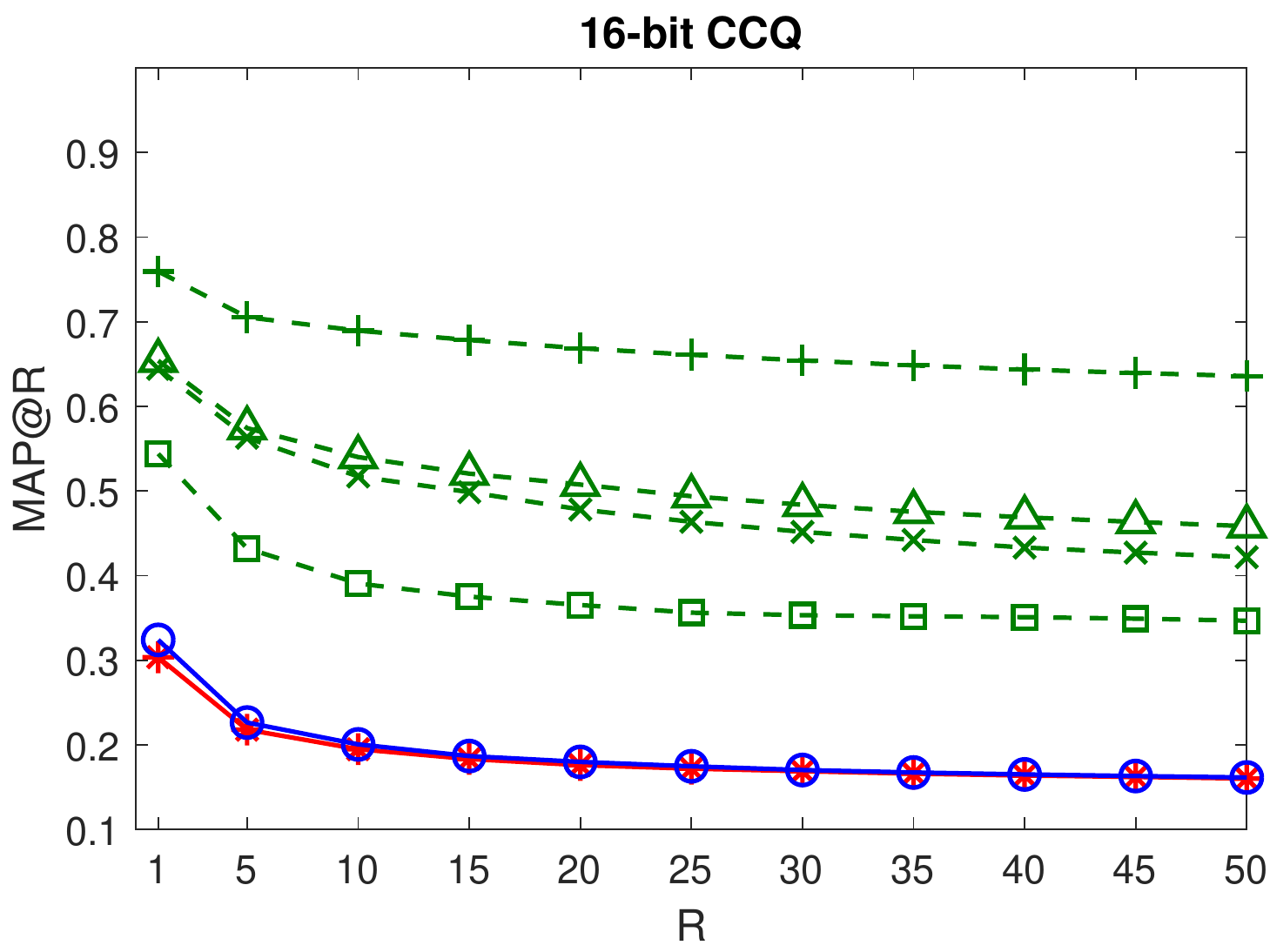}
    \\
    \includegraphics[height=.15\textheight,width=.25\textwidth]{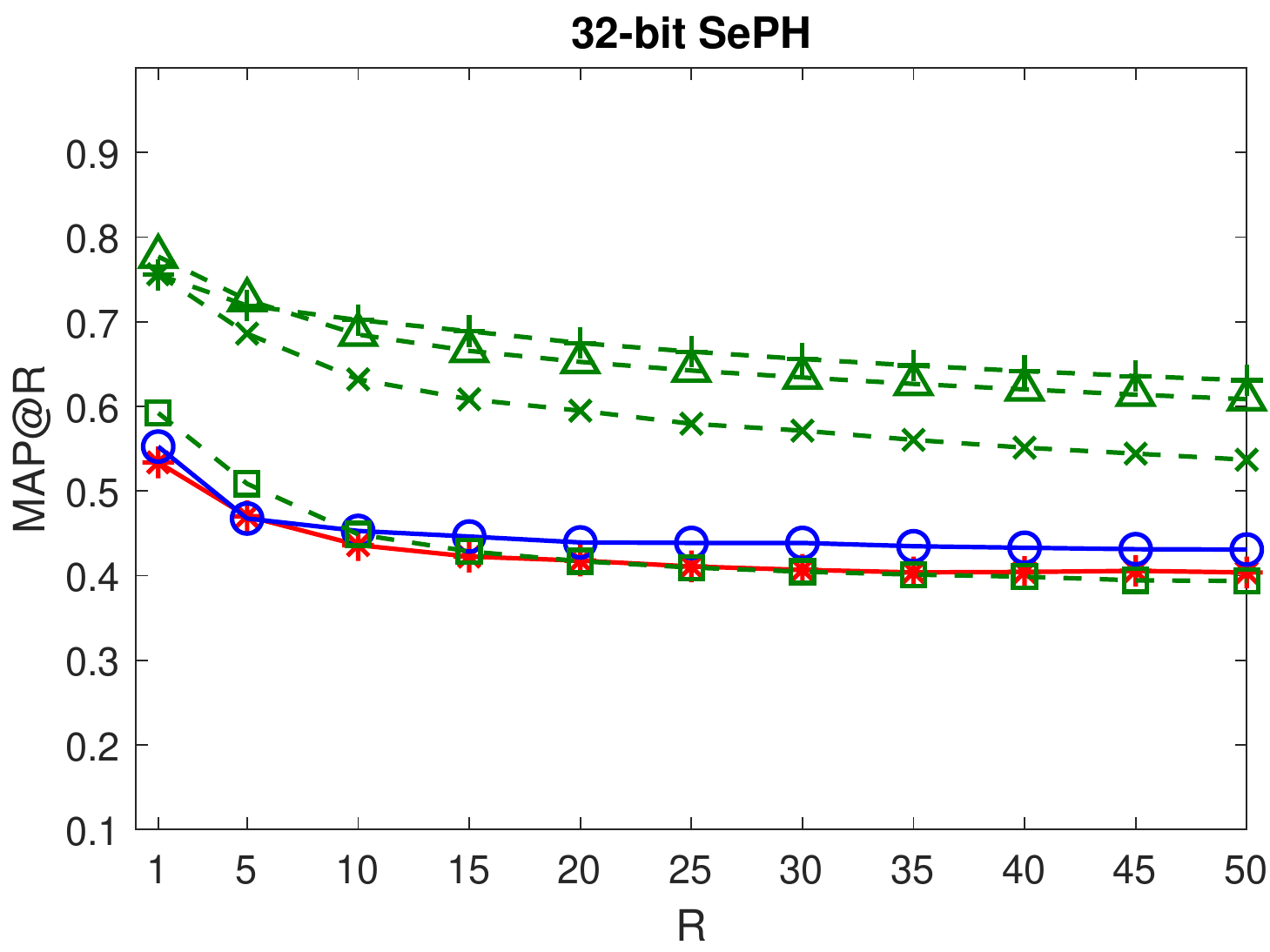} 
    \includegraphics[height=.15\textheight,width=.25\textwidth]{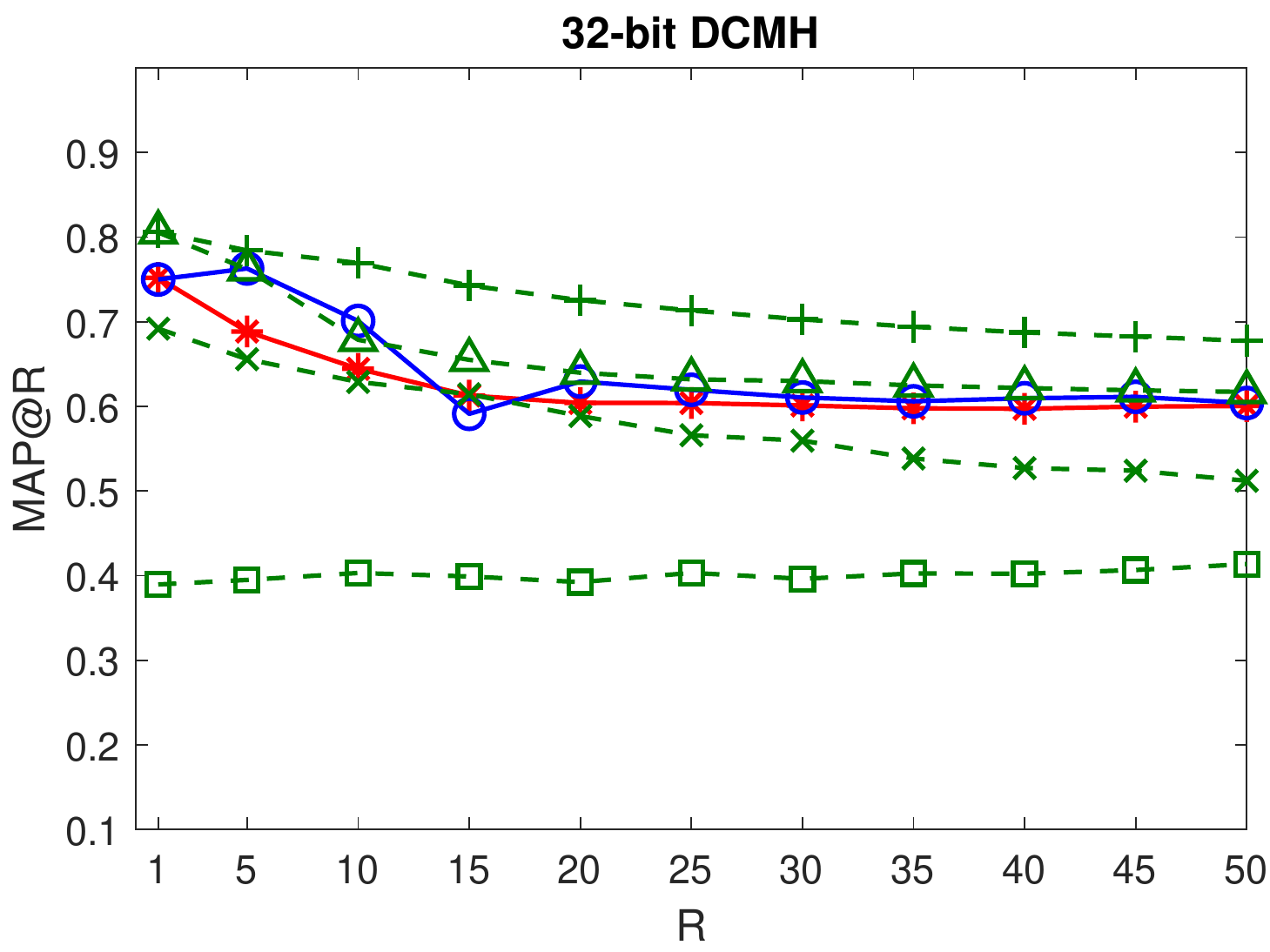} 
    \includegraphics[height=.15\textheight,width=.25\textwidth]{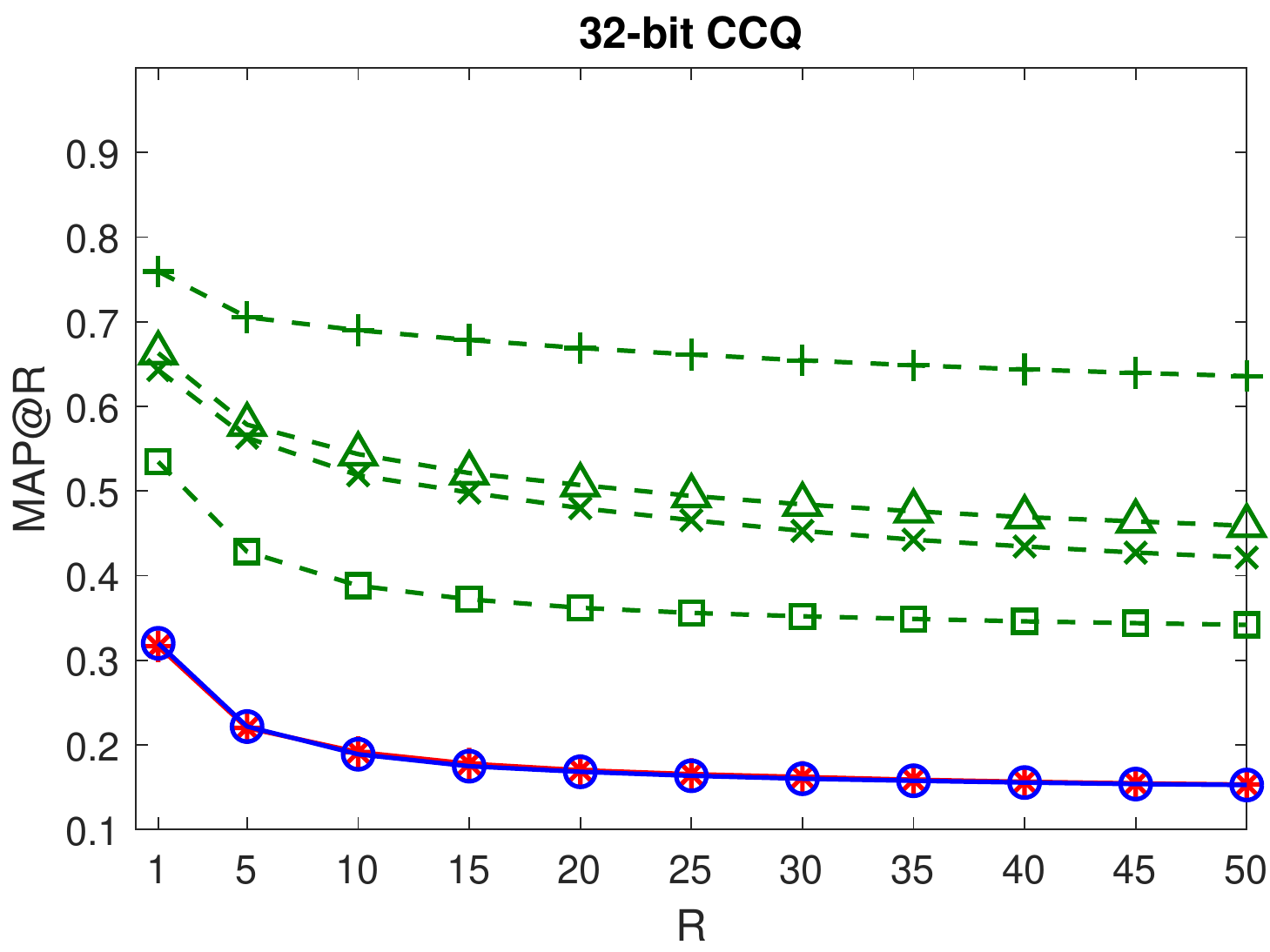} 
  \end{tabular}  
  \caption{MAP@R in the NUS-WIDE dataset for image query vs. text dataset.}
  \label{fig:32_it}
\end{figure*}

Figures \ref{fig:16_ti} and \ref{fig:32_ti} show the results for the search modality ``text query vs. image dataset" ($T \rightarrow I$) in MIRFlickr and NUS-WIDE datasets, respectively, and Figures \ref{fig:16_it} and \ref{fig:32_it} demonstrate another search modality ``image query vs. text dataset" ($I \rightarrow T$).
The X-axis and Y-axis represent the number of retrieved examples \textit{R} and MAP@\textit{R}, respectively. 
Except for the MIRFlickr-CCQ dataset, the exhaustive and naïve index schemes have similar MAP curves.
The former scheme did not benefit from reranking all reference data since their Hamming distances do not accurate enough to reflect the similarities to the query. 
However, the latter scheme can reach a comparable accuracy by taking only a few candidates for reranking.
Moreover, with the proposed DNN-guided index scheme, we can effectively boost the accuracy compared with the above baseline schemes.
We observe that the longer index code performed stably and yielded close MAP curves across various CMH algorithms.
In addition, the longer index code generated a more compact candidate list.
Tables \ref{tab:comparisonT2I} and \ref{tab:comparisonI2T} compare our method with these CMH algorithms for $T \rightarrow I$ and $I \rightarrow T$, respectively, in terms of MAP@50, the fraction of accessed reference data (ARD\%), and runtime. ARD\% is defined by:
\begin{equation}
	\mathrm{ARD\%} = \frac{\text{the number of candidates}}{\text{the number of reference data points}} \times 100\%.
\end{equation}

A lower ARD\% means a smaller computation cost due to less memory access operations for the reference data.
The 14-bit DNN-index scheme, which obtained the highest accuracy and smallest computation cost, showed a significant improvement when it integrated with these CMH methods.

\begin{table*}[!htb]
\centering
  \caption{Comparison in terms of MAP@50, ARD\%, and runtime (milliseconds per query) for text query vs. image dataset}
  \label{tab:comparisonT2I}
  \begin{tabular}{|l||c|c|c|c|c|c||c|c|c|c|c|c|}
  \hline
  \multirow{3}{*}{} & \multicolumn{6}{c||}{MIRFlickr}                         & \multicolumn{6}{c|}{NUS-WIDE}                         \\ \cline{2-13} $T \rightarrow I$ 
                  & \multicolumn{3}{c|}{16-bit} & \multicolumn{3}{c||}{32-bit} &  \multicolumn{3}{c|}{16-bit} & \multicolumn{3}{c|}{32-bit} \\ \cline{2-13} 
                  &  MAP@50  &  ARD\% &  time  &  MAP@50  &  ARD\% &  time   &  MAP@50  &  ARD\%  & time  &  MAP@50  &  ARD\% &  time  \\ \hline
    SePH  & 0.7137 & 100\% & 1.40 & 0.7493 & 100\% & 2.07 & 0.5303 & 100\% & 13.33 & 0.5992 & 100\% & 20.83\\ \hline
    DCMH  & 0.7451 & 100\% & 1.49 & 0.7660 & 100\% & 2.19 & 0.5777 & 100\% & 13.20 & 0.5961 & 100\% & 19.80\\ \hline
    CCQ   & 0.4842 & 100\% & 1.70 & 0.4539 & 100\% & 3.04 & 0.1666 & 100\% & 13.85 & 0.1565 & 100\% & 22.35\\ \hline
    DNN-index (14 bits)  & \textbf{0.8753} & \textbf{0.33\%} & \textbf{1.04} & \textbf{0.8754} & \textbf{0.33\%} & \textbf{1.04} & \textbf{0.7564} & \textbf{0.23\%} & \textbf{1.27} & \textbf{0.7583} & \textbf{0.23\%} & \textbf{1.27} \\ \hline
  \end{tabular}
\end{table*}

\begin{table*}[!htb]
\centering
  \caption{Comparison in terms of MAP@50, ARD\%, and runtime (milliseconds per query) for image query vs. text dataset}
  \label{tab:comparisonI2T}
  \begin{tabular}{|l||c|c|c|c|c|c||c|c|c|c|c|c|}
  \hline
  \multirow{3}{*}{} & \multicolumn{6}{c||}{MIRFlickr}                         & \multicolumn{6}{c|}{NUS-WIDE}                         \\ \cline{2-13} $I \rightarrow T$ 
                  & \multicolumn{3}{c|}{16-bit} & \multicolumn{3}{c||}{32-bit} &  \multicolumn{3}{c|}{16-bit} & \multicolumn{3}{c|}{32-bit} \\ \cline{2-13} 
                  &  MAP@50  &  ARD\% &  time  &  MAP@50  &  ARD\% &  time   &  MAP@50  &  ARD\%  & time  &  MAP@50  &  ARD\% &  time  \\ \hline
    SePH  & 0.5992 & 100\% & 1.39 & 0.6179 & 100\% & 2.10 & 0.3747 & 100\% & 13.29 & 0.4037 & 100\% & 20.54\\ \hline
    DCMH  & 0.6899 & 100\% & 1.54 & 0.7075 & 100\% & 2.44 & 0.4823 & 100\% & 14.97 & 0.6005 & 100\% & 19.34\\ \hline
    CCQ   & 0.4011 & 100\% & 1.77 & 0.3996 & 100\% & 2.94 & 0.1601 & 100\% & 13.92 & 0.1530 & 100\% & 22.97\\ \hline
    DNN-index (14 bits)  & \textbf{0.8803} & \textbf{0.32\%} & \textbf{0.93} & \textbf{0.8803} & \textbf{0.32\%} & \textbf{0.93} & \textbf{0.6775} & \textbf{0.03\%} & \textbf{1.02} & \textbf{0.6775} & \textbf{0.03\%}  & \textbf{1.02} \\ \hline
  \end{tabular}
\end{table*}

\section{Conclusion}

In this paper, we propose a novel search method that utilizes a probability-based index scheme over binary hash codes in cross-modal retrieval.
The index scheme, which ranks the hash index codes of the inverted table through DNN, can effectively increase the search accuracy and decrease the computation cost.
Extensive experimental results show the superiority of the proposed method compared with other baselines.

\section*{Acknowledgement}

This work was supported by the Ministry of Science and Technology, Taiwan, under grants MOST 106-2221-E-415-019-MY3.

\end{document}